\documentclass{pasa}
 \usepackage{tabularx}
 \usepackage{ragged2e}
\usepackage[T1]{fontenc}
\usepackage[latin9]{inputenc}
\usepackage{fancyhdr}
\usepackage{array}
\usepackage[authoryear]{natbib}

\newcommand{\msun}{M$_\odot$}
\newcommand{\rsun}{R$_\odot$}

\newcommand{\mytilde}{\raise.17ex\hbox{$\scriptstyle\mathtt{\sim}$}}
\def\lesssim{\mathrel{\hbox{\rlap{\hbox{\lower3pt\hbox{$\sim$}}}\hbox{\raise2pt\hbox{$<$}}}}}
\def\gtrsim{\mathrel{\hbox{\rlap{\hbox{\lower3pt\hbox{$\sim$}}}\hbox{\raise2pt\hbox{$>$}}}}}
\def\lesseq{\mathrel{\hbox{\rlap{\hbox{\lower3pt\hbox{$-$}}}\hbox{\raise2pt\hbox{$<$}}}}}
\def\gtreq{\mathrel{\hbox{\rlap{\hbox{\lower3pt\hbox{$-$}}}\hbox{\raise2pt\hbox{$>$}}}}}

\title[The impact of companions on stellar evolution]{The impact of companions on stellar evolution }
\author[De Marco \& Izzard]{Orsola De Marco$^{1,2}$
\& \rgi{Robert}~G. Izzard$^3$\\
\affil{$^1$Department of Physics \& Astronomy, Macquarie University, Sydney, NSW 2109, Australia}
\affil{$^2$Astronomy, Astrophysics and Astrophotonics Research Centre, Macquarie University, Sydney, NSW 2109, Australia}
\affil{$^3$Institute of Astronomy, University of Cambridge, Cambridge, \rgi{CB3 0HA, United Kingdom.}}
}

\jid{PASA}
\doi{10.1017/pas.\the\year.xxx}
\jyear{\the\year}

\usepackage[authoryear]{natbib}
\bibpunct{(}{)}{;}{a}{}{,}
\usepackage{xcolor}
\usepackage{ulem}

\newcommand\rgi[1]{#1}
\newcommand\rgix[1]{}

\begin{document}

\begin{abstract}
Stellar astrophysicists are increasingly taking into account the effects of orbiting companions on stellar evolution. New discoveries, many thanks to systematic time-domain surveys, have underlined the role of binary star interactions in a range of astrophysical events, including some that were previously interpreted as due {\it uniquely} to single stellar evolution. Here, we review classical binary phenomena such as type Ia supernovae, and discuss new phenomena such as intermediate luminosity transients, gravitational wave-producing double black holes, or the interaction between stars and their planets. Finally, we examine the reassessment of well-known phenomena in light of interpretations that include both single {\it and} binary stars, for example supernovae of type Ib and Ic or luminous blue variables. At the same time we contextualise the new discoveries within the framework and nomenclature of the corpus of knowledge on binary stellar evolution.
 The last decade has heralded an era of revival in stellar astrophysics as the complexity of stellar observations is increasingly interpreted with an interplay of single and binary scenarios. The next decade, with the advent of massive projects such as the {\it Large Synoptic Survey Telescope}, the {\it Square Kilometre Array}, the {\it James Webb Space Telescope}  and increasingly sophisticated computational methods, will see the birth of an expanded framework of stellar evolution that will have repercussions in many other areas of astrophysics such as galactic evolution and nucleosynthesis.
\end{abstract}


\maketitle

\section{INTRODUCTION }

\label{sec:intro}

Many classes of stars are either known or presumed to be binaries and many astrophysical observations can be explained by interactions in binary or multiple star systems. This said, until recently duplicity was not widely considered in stellar evolution, except to explain certain types of phenomena. 

Today, the field of stellar astrophysics is fast evolving. Primarily, time-domain surveys have revealed a plethora of astrophysical events many of which can be reasonably ascribed to binary interactions. These surveys reveal the complexity of binary interactions and also provide sufficient number of high quality observations to sample such diversity. In this way, new events can be classified and some can be related to well known phenomena for which a satisfactory explanation had never been found. Binary-star phenomena are thus linked to outstanding questions in stellar evolution, such as the nature of type Ia, Ib and Ic supernovae, and hence are of great importance in several areas astrophysics.

Another \rgi{recent and important discovery} is that most main-sequence, massive stars are in multiple systems, with $\sim$70 per cent of them predicted to interact with their companion(s) during their lifetime (e.g., \citealt{Sana2012}, \citealt{Kiminki2012}, \citealt{Kobulnicky2012}, \citealt{Kobulnicky2014}). 
This \rgi{discovery suggests} that many massive star phenomena are related to the presence of a binary companion, \rgix{such as,} for example, \rgi{type Ib and Ic} supernova\rgi{e} \citep{Podsiadlowski1992,Smith2011b} or \rgix{even classic} phenomena such as luminous blue variables \citep{Smith2015}. Additionally, the recent detection of gravitational waves has confirmed the existence of binary black holes \citep{Abbott2016}, which are a likely end-product of the most massive \rgix{stars,} binary evolution.

Finally, we now know that planets exist frequently around \rgi{main-sequence} stars. 
\rgi{Planets have also been} discovered orbiting evolved stars, close enough that an interaction must have taken place. 
These discoveries open the possibility that interactions between stars and planets may change not only the evolution of the planet or planetary system, but the evolution of the mother star. 

The fast\rgi{-}growing corpus of binary interaction observations also allows us to conduct experiments with stellar structure, because companions perturb the star. Examples are the "heartbeat stars", which are eccentric binaries with intermediate orbital separation. At periastron the star is "plucked" like a guitar string and the resulting oscillation spectrum, today studied thanks to high precision observations such as those of {\it Kepler Space Telescope}, can be used to study the stellar layers below the photosphere \citep{Welsh2011}.

\begin{figure}
\centering
\includegraphics[width=0.5\textwidth]{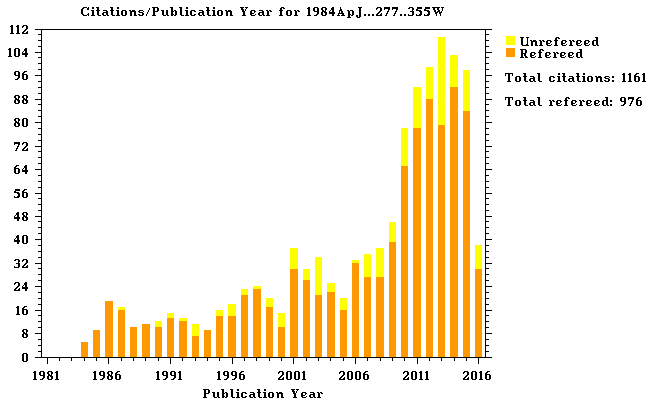}
\caption{{ Citations to the seminal paper on the common envelope binary interactions of Webbink (1984). The relative increase starting  in approximately 2006 cannot be explained by the overall increase in the number of astrophysics papers over the same period, demonstrating an increase in interest in this interaction over the last 10 years. {\it Figure sourced from the Astrophysics Data Service}}.}
\label{fig:citations}
\end{figure}

Alongside new observations, creative new methods exist to model binaries. 
One-dimensional stellar structure and evolution codes such as the new {\it Modules for Experiments in Stellar Astrophysics} ({\sc mesa}) are used to model not just stellar evolution, but binary processes such as accretion. 
\rgi{Multi-dimensional simulations able to model accurately both the stars and the interaction are currently impossible because of the challenge in modelling simultaneously a large range of space and time scales.} 
However, great progress over recent years has been made in 3D hydrodynamics with 3D models of individual stars (e.g., \citealt{Chiavassa2011}, where important relevant physics is captured) and numerical techniques being refined and developed to be adopted for binary modelling \citep[e.g.,][]{Ohlmann2016,Eliot2016}. 
These studies are part of a revival in the field of stellar astrophysics, the start of which may be observed in the increase by over a factor of 10 in citations to seminal binary interaction papers such as that of \citet[][Fig.~\ref{fig:citations}]{Webbink1984}, well above the factor of 2.3 increase in the overall volume of astrophysics papers that has been witnessed between 1985 and 2015. 

This review is \rgi{arranged} as follows\rgi{. W}e start with a \rgix{very short} summary of the properties of main sequence binary populations (Section~\ref{sec:main-sequence-binary-stars}), including stars with planetary systems (Section~\ref{ssec:star-planet-systems}). In Section~\ref{sec:binary-classes} we briefly discuss binary pathways and list binary classes. In Section~\ref{sec:the-binary-star-toolkit} we discuss current and future observational platforms particularly suited to the study of binary phenomena as well as modelling tools used to interpret observations and make predictions. We then, in Section~\ref{sec:binaries-as-laboratories}, emphasise how certain classes of binaries allow us to carry out stellar experiments, while in Section~\ref{sec:interesting-and-curious-binary-phenomena} we report a range of interesting phenomena, which have been explained by including the effects of interactions between stars or between stars and planets. In Section~\ref{sec:tran!
 sients} we discuss the \rgi{exciting and expanding} field of stellar transients, including the newly detected gravitational waves. We conclude in Section~\ref{sec:concluding-remarks}. 

\section{MAIN-SEQUENCE BINARY STARS}
\label{sec:main-sequence-binary-stars}

In this section we summarise the frequencies of main sequence binaries as well as their period and mass ratio distributions. For a recent review of stellar multiplicity in pre-main sequence and main sequence stars see \citet{Duchene2013}. 

The fraction of stars that \rgi{interact} with their companions depends on these \rgi{frequencies} and on the action of tides \rgi{and mass loss}, which can \rgi{both} shorten \rgi{and lengthen} the orbital separation, bringing two stars within each other's influence or allowing them to avoid an interaction altogether. Here, \rgi{we define} the binary fraction \rgi{to be} the fraction of \rgi{systems that are multiple} rather than the companion frequency, \rgi{which} can be larger than unity \rgi{when} there is more than one companion per primary on average. 

We use the same naming conventions as \citet{Duchene2013}\rgi{. Massive stars are more massive than about 8~\msun, intermediate-mass stars have masses from about 1.5~to~5~\msun\ (spectral types B5~to~F2), Solar-type stars have masses in the approximate range 0.7~to~1.3~\msun\ (spectral types F~through~mid-K), low-mass stars between 0.1 and 0.5~\msun\ (spectral types M0~to~M6) and very-low-mass stars and brown~dwarfs are less massive than about 0.1~\msun\ (spectral types M8~and~later). }

\subsection{The multiplicity fraction of main sequence stars}
\label{ssec:the-binary-fraction}

Our knowledge of the fraction of massive stars that have companions was considerably revised by the Galactic O-star survey of \citet{Sana2012} and \citet{Kiminki2012}. \citet{Sana2012} show that the fraction of O stars that have companions that will interact during the lifetime of the O star\rgi{, i.e.~with }periods \rgi{shorter} than \rgi{about} $1500$~days\rgi{,} is 71 per cent. Of these, one third \rgix{are thought to }merge. The overall binary fraction is established to be \rgi{more than 60 per cent in} early B stars \rgi{and in excess of} 80 per cent \rgi{in} O stars. There is evidence that the binary fraction in clusters and in the field are similar \citep{Duchene2013}.

The fraction of intermediate-mass stars in binary systems is substantially lower than for the most massive stars.  
\rgi{Amongst} the entire group of \rgi{F2~to~B5} type stars, the binary frequency is greater than about 50 per cent, \rgi{as} determined from \rgix{analysing} the Sco-Cen OB association \citep{Kouwenhoven2005,Fuhrmann2012,Fuhrmann2015}. 

As shown by \citet{Duquennoy1991} and, later, by \citet{Raghavan2010}, about $44\pm2$ per cent of G-~and~K-type main sequence stars in the Solar neighbourhood have companions\rgi{. They} point out a low-significance difference of 9 percent points between the immediately sub-Solar $41\pm3$ per cent (G2-K3) and immediately super-Solar, $50\pm4$ per cent (F6-G2) primaries. 

The binary fraction\rgi{s} of low\rgi{-} and very\rgi{-}low mass stars \rgi{are} $26\pm3$ per cent and $22 \pm 5$ per cent, respectively \citep{Duchene2013}. These stars are not massive enough to evolve \rgi{off the main sequence} within the age of the \rgi{Universe. They} are \rgi{thus} more interesting \rgi{as} companions to more massive stars\rgix{, rather} than \rgi{as} primaries in \rgi{all but the closest binaries}. 

\subsection{The period distribution of main sequence binaries}
\label{ssec:the-period-distribution}

The initial period distribution\footnote{By "initial" we mean the zero-age main sequence. Orbital elements are altered during the pre-main sequence phase as well as during the main sequence itself. Such changes are interesting in themselves as they pertain to the method of binary formation and the action of tides during the main sequence.} of massive binaries has a peak at very short periods and declining numbers of larger period binaries out to a period of approximately 3000 days. Two separate distributions are envisaged, a population of short period binaries (periods shorter than approximately one day) and a slowly declining power-law period distribution extending out to 10\,000~AU \citep{Sana2012b}.

In intermediate mass stars of spectral type A and B, the initial period distribution lies between that of massive stars (Sana et al. 2011), which is a power law, and the longer period, log-normal distributions of Solar-like stars \citep[][see below]{Duquennoy1991,Raghavan2010}. Indeed, \citet{Kouwenhoven2005,Kouwenhoven2007} found that the observed intermediate mass binaries fit equally well a distribution flat in log-period (\"Opik's distribution). In fact, alone among all binaries, the A-type stars appear to present a double peaked period distribution with a peak below 1~AU and one at about 350~AU \citep{Kouwenhoven2007,Duchene2013}, although there is a great deal of uncertainty in these statements.

In Solar-like stars, the distribution is log-normal, peaked at about $10^4\, \mathrm{days}$ (Raghavan et al. 2010). This means that relatively few Solar-like binaries interact with their companions compared to more massive stars that are not only in binaries more often, but that have systematically closer companions.

Additional and updated information on the period distribution can be found in \citet{Moe2016} who have reanalysed all existing binary observations.

\subsection{The mass ratio distribution of main sequence binaries}
\label{ssec:mass-ratio-distribution}

The more massive the companion relative to the primary, the \rgi{greater} the impact on the primary star once the binary interacts. Hence the steeper the exponent, $\gamma$, of the mass ratio distribution \rgi{($f(q) = q^\gamma$ where $q=M_2/M_1$ and $M_{1,2}$ are the most and least massive component of the binary, respectively)}, the more dramatic the interactions in that population of binaries will be. 

Equal mass binaries are not favoured in the massive star population \rgi{\citep{Sana2012}}, nor is there any  evidence that the mass ratio distribution is different in wider and closer binaries ($\gamma = -0.1 \pm 0.6$ \rgi{at} $\log P \lesseq 3.5$ and $q \gtreq 0.1$ and $\gamma = -0.55 \pm 0.13$ \rgi{when} $a \gtreq 100$~AU). A \rgi{peak in $f(q)$} at $q \approx$0.8 does not point to a separate population of "twins" \citep{Lucy1979}. A small population of high mass ratio binaries could be due to mass-transfer during the pre-main sequence phases. 

Intermediate mass stars \rgix{seem to} have a shallow mass-ratio distribution ($\gamma=-0.45 \pm 0.15$) \rgi{in} both compact and wider binaries, although incompleteness remains a problem. Earlier claims that high mass-ratio systems were favoured also in intermediate-mass stars, no longer seem to hold \citep{Duchene2013}. 

The distribution of mass ratios in Solar-type stars is approximately flat (Raghavan et al. 2010). \citet{Duchene2013} re-fitted the data of \citet{Raghavan2010} and found that longer period binaries \rgi{($\log P/\mathrm{days} > 5.5$)} have a flat mass-ratio distribution ($\gamma=-0.01 \pm 0.03$), while closer binaries \rgi{($\log P/\mathrm{days} \lesseq 5.5$)} have a higher incidence of components with similar masses ($\gamma = 1.16 \pm 0.16$). There is a remarkable lack of substellar companions around Solar-type stars (the brown dwarf desert; \citealt{Marcy2000,Grether2006}).

\rgi{The situation is similar \rgi{in} low mass stars. 
The mass ratio distribution is flat \rgi{among} the wide binaries, with $\gamma = -0.2 \pm 0.3$ among systems wider than 5~AU \citep{Duchene2013}.
Binaries closer than 5~AU have $\gamma = 1.9 \pm 1.7$ as determined by fitting the data of the RECONS consortium \citep{Henry2006}.}
Brown\rgi{-}dwarf companions are easier to find around low\rgi{-}mass stars than around Solar-type stars. When brown dwarf\rgi{s} are the primaries in binaries, their companions tend to \rgi{be of similar} mass, \rgi{suggesting a large $\gamma$} \citep{Burgasser2006}.
 
Universally, multiplicity surveys are \rgi{incomplete when} $q\lesssim0.1$. 
As a result, \rgi{in binaries with primary masses around} 1~\msun, the limit for \rgi{detection of a companion} is \rgix{approximately} at the star-brown dwarf boundary. With only few brown dwarfs to be found around such stars, this bias may not result in a substantial underestimation of the number of companions around Solar-like stars. 
\rgi{However, amongst stars more massive than 10~\msun, the lack of information on binaries with $q\lesssim0.1$ means that the frequency of companions with mass as high as approximately 1~\msun~is unknown}. If an interaction with a much less massive companion \rgi{results} in a detectable change in the primary's evolution, then we would like to know their numbers so as to account correctly for the frequency of these interactions.

\citet{Moe2015} \rgi{describe} a handful of B\rgi{-}type stars with a very low mass, pre-main sequence companion. From these discoveries they infer that B-type binaries with extreme mass ratios ($q<0.25$) are in binaries only one third of the times of B-type binaries with more comparable masses ($q>0.25$) for the same period range. They also \rgi{conclude} that the frequency of close, low mass companions is a strong function of primary mass. This would \rgi{render} these low\rgi{-}mass companions \rgi{un}important to the binary evolution of Solar-type stars, but perhaps very important for more massive primaries as there are more of them. 

Additional and updated information on the mass ratio distribution can be found in \citet{Moe2016} who have reanalysed all existing binary observations.

This question can be extended to whether the interaction with a planetary mass companion would have any effect on the star. For example, would an interaction with Jupiter affect the evolution of the Sun? Planets are present around a substantial fraction of all main sequence stars. If these planets do alter the evolution of their mother stars, then such interaction must be taken into account in stellar evolution (Section~\ref{ssec:star-planet-systems}).

\subsection{Orbital eccentricity of main sequence binaries}
\label{ssec:orbital-eccentricity-of-main-sequence-binaries}

Eccentricity is an important parameter in intermediate-mass systems. Recent studies, e.g.,~\citet{Tokovinin2015}, 
suggest the number of stars, $N$, with a given eccentricity, $e$, is $dN/de \approx 1.2 e + 0.4$, with a slight dependence on orbital separation. Close binaries probably tidally circularise prior to mass transfer by Roche-lobe overflow \citep[][see Section~\ref{sec:binary-classes}]{Hurley2002}, but intermediate-period, eccentric binaries may undergo episodic mass transfer, which is as yet poorly understood \citep{Sepinsky2009,Lajoie2011b}.  We discuss tidal circularisation further in Section~\ref{ssec:wind-polluted-stars}.

The distribution of the orbital eccentricity does not appear to depend on \rgi{the mass of a binary system}. Interestingly\rgi{, it seems to be distributed similarly when}  companion is a brown dwarf or a planet, showing that it is likely imparted by dynamical processes rather than star formation. 
In Section~\ref{ssec:main-sequence-triples-and-higher-order-multiple-stars} we \rgix{will} dwell further on orbital eccentricity because changes in its value at the hand of a second, wide companion can bring an inner companion, originally in an orbit too wide for interaction, into contact with the evolving primary.





%




\subsection{Star-planet systems}
\label{ssec:star-planet-systems}

Planets around main sequence stars \rgix{will} interact with the star if the orbital separation decreases or if the star expands to fill its Roche lobe (Section~\ref{sec:binary-classes}), provided that mass loss does not first widen the planetary orbit, reducing any chance of Roche lobe overflow. 

It is not clear how these interactions \rgix{will} alter the evolution of the stars \citep{Soker1998d,Nelemans1998,Nordhaus2006,Staff2016b}. Effects of star-planet interactions may include spinning up of the star \citep[e.g.,][]{Carlberg2009}, pollution of the stellar envelope with materials from the planets \citep{Sandquist2002}, increase of the mass-loss rate of the star in response to an interaction with the planet, or with its gravitational potential \citep{Bear2011}. 

No planet has \rgix{ever} been observed plunging into a star \rgi{even though} such interactions must \rgi{occur}. If planets plunge into their star during \rgi{its} main sequence phase, it is possible that only limited changes would be observed in the star. The most likely effect is the spin-up of the giant's envelope \citep{Privitera2016b} while the enhancement of certain elements such as carbon, iron or lithium is either modest of difficult to interpret due to various competing effects \citep{Dotter2003,Privitera2016b,Staff2016b}. However, \rgi{if particularly massive planets plunge} into their star \rgi{after the main sequence}, when the stellar envelope is more extended and less \rgi{gravitationally} bound, it is possible that other effects may become observable. To plunge into the star when the star is extended, the planet needs to be at initial orbital separation between 2 and 4 times the radius the star has during its giant phases (depending on \rgi{the strength of t!
 ides}; e.g., \citealt{Villaver2009}, \citealt{Mustill2012}, citealt{Madappatt2016}). Unfortunately, techniques such as radial velocity variability or microlensing, which have yielded many planets \citep{Sousa2011}, only detect planets out to a few \rgi{AU. 
Imaging} surveys \rgi{able} to detect high \rgi{contrast are} few and have detected only a handful of planets out to tens of AU \citep[][]{Soummer2011,Kuzuhara2013}. 
A handful of planets are also \rgi{known} around giant stars \citep[e.g.,][]{Niedzielski2015}, \rgi{although} these surveys are incomplete. As a result there is no way, yet, to know how often planets orbit stars at distances such that the system will interact during the giant phases of the star.

There is some evidence that star-planet interactions may have taken place. For example, planets have been discovered in \rgi{close} orbits around stars whose precursor was larger than today's orbit. \rgi{Examples are} the two earth-mass planets around KIC~05807616 \citep{Charpinet2011}, which may be the disrupted cores of more massive gas giants \citep{Passy2012b}. All of these discoveries are indirect and there is a suspicion that at least some of them may be due to phenomena unrelated to the presence of a planet. A particular example is that of planets discovered around binaries that are known to have gone through a common envelope interaction (see Section~\ref{sec:binary-classes}). In most cases the (indirect) detections of such planets have either been dismissed \cite[][]{Parsons2010b} or the deduced planets' orbits have been shown to be dynamically unstable, indicating that the observations are more likely to need an alternative explanation \cite[][]{Potter2011,Hinse201!
 2,Horner2013}. In other cases it appears that the planetary interpretation is the only one \citep[e.g., NN~Ser;][]{Parsons2014}, but that the planet was formed {\it as a result} of the binary interaction rather than at the time of star formation \citep{Beuermann2011,Veras2012}. Second generation planet formation remains an unexplored area of Astrophysics \citep{Baer2014}, although we know that it must take place because of the presence of planets around some pulsars \citep{Wolszczan1992}.

\subsection{Main-sequence triples and higher order \\ multiple stars}
\label{ssec:main-sequence-triples-and-higher-order-multiple-stars}

The fraction of systems with a tertiary companion can be gauged by subtracting the fraction of multiples from the fraction of companions, while assuming that\rgix{, in first approximation,} multiple systems comprise only binaries and triple systems. \rgi{In} table~1 of \citet{Duchene2013} we see that the \rgi{frequency} of triples increases dramatically with \rgi{primary mass} from \rgi{0 per cent in late-M} stars and brown dwarfs to \rgi{about half for} stars more massive than 1.5~\msun. Also, the closest binaries have tertiary companions far more often than wider \rgi{binaries. Solar-type stars in binaries with periods shorter than 3~days have tertiaries in 96 per cent of systems, compared to only 34 per cent in binaries with periods longer than 12 days \citep{Tokovinin2006}.}

Such tertiary companions are of interest to stellar evolution because they may act to shorten the orbital period of the inner binary and alter the inner binary eccentricity. This can bring an inner binary initially too wide for an interaction during stellar evolution to within the reach of the expanding primary, therefore increasing the number of interactions in a population. For example \citet{Hamers2013} find that in 24 per cent of the triple systems they studied with a population synthesis technique, the inner binary, which initially is wide enough to avoid interaction, is hardened enough for mass transfer to start.

Finally, one of the most important reasons why we need to model mass transfer correctly is because it has an impact on the orbital elements, which in turn decide the strength and type of the interaction that follows. While analytical work has moved this field a long way \citep[e.g.,][]{Sepinsky2007b,Sepinsky2009,Sepinsky2010,Dosopoulou2016,Dosopoulou2016b} it is not clear how to implement this information in numerical codes \citep{Staff2016,Iaconi2016}. There, small stellar deformation or oscillations due to numerical artefacts translate into strong tidal torques that alter the orbit giving reasonable doubt that the orbital evolution (and hence the interaction) is well reproduced.

\subsection{Binaries born in cluster environments}
\label{ssec:binaries in cluster environments}

One of the fundamental questions of binary star formation is whether the binary frequency, period, mass ratio and orbital eccentricity distributions are dependent on the birth environment. An obvious check is to compare these quantities for binaries in different clusters and in the field. 

This is complicated by the fact that cluster environments, with a range of stellar densities, may alter the binary fraction and other characteristics of the binary population over time in different ways. The binary population thus carries the signature of its environment of birth, but also of the cluster's age.  Another caveat is that binaries in the field were born in different clusters that dissipated at different times. Hence their binary characteristics would be a mix. Clearly comparing binary populations in clusters and in the field is not easy, as there are many variables that contribute to their appearance, but this should not deter us from trying!

Recently, a thorough study of the binary population in M~35, a young, 180 Myr old cluster \citep{Geller2010}, has revealed that the binary properties that could be probed by their study are similar to those of the binaries known in the field. More specifically, binaries and single stars show no difference in distribution within the cluster, and the binaries are not centrally concentrated. The binary frequency out to period of $10^4$~days is $24\pm3$ per cent, consistent with the binary frequency in the field to the same period limits \citep{Duquennoy1991}. This would argue that the field binary population is similar to that of a cluster that did not have time to alter the characteristics of its binaries. It seems also to argue that field binaries  were not, on average, altered significantly while they inhabited their birth clusters. 

In Section~\ref{ssec:binary-stars-that-evolve-in-clusters} we discuss how cluster dynamics change the binary population and contribute exotic binary formation.

\section{BINARY EVOLUTION AND CLASSES: \\ PATHWAYS AND NOMENCLATURE}
\label{sec:binary-classes}

\rgi{A} multiple star system \rgi{has} as many sets of stellar parameters \rgi{-- mass, metallicity, etc. --} as there are stars in the system. 
\rgi{In addition there are} parameters characterising the orbital elements, such as separation, eccentricity, spin alignment \citep{Hut1981}, or orbital alignment \rgi{in} multiple systems. 
This \rgi{large parameter space introduces} complexity, particularly when the two or more stars interact, and two relatively similar systems \rgi{may} display quite diverse phenomenology.

Below we summarise the types of binary interactions. We then compile a list of binary class names. Finally we discuss how binary interactions play out in clusters, where encounters between stars and binaries are frequent.

\subsection{Types of binary mass transfer}
\label{ssec:types-of-binary-mass-transfer}

\begin{figure}
\centering
\includegraphics[width=0.5\textwidth]{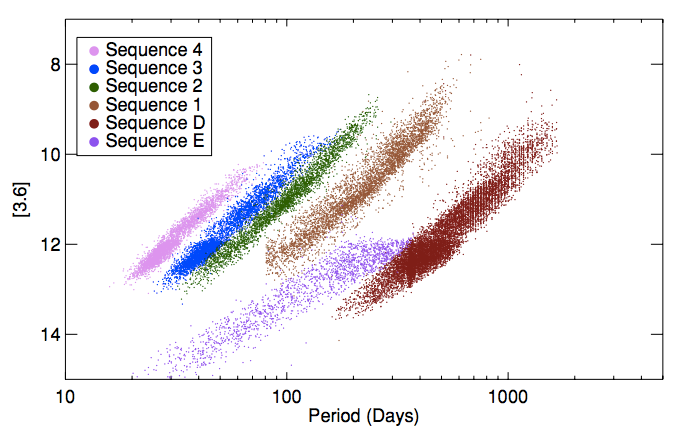}
\caption{Period-luminosity diagram for evolved stars in the Large Magellanic Cloud. Sequence
1 consists of stars pulsating in the fundamental mode, while sequences 2 to 4
are higher order pulsational modes. Sequence E (Section~\ref{ssec:types-of-binary-mass-transfer}; Table~1) stars are ellipsoidal binary
systems, where a close companion distorts the giant primary star. 
The mechanism responsible for the variation on sequence D is not
known. {\it Credit: image adapted from figure 1 of \citet{Riebel2010}}.}
\label{fig:sequenceE}
\end{figure}

Binary classes can be best understood by thinking about the possible range of interactions that take place between two stars. The mass stored in a star is layered on equipotential surfaces. Stars remain spherical when they are small and compact, but as they age and expand their surfaces become distorted by the gravity of their companion star and by rotation which can be enhanced by tidal interactions with a companion. Distorted stars can sometimes be observed as "ellipsoidal variables" (e.g., the sequence E stars; \citealt{Nicholls2012}, see Table~1 and Fig.~\ref{fig:sequenceE}).

A special equipotential surface, called the Roche surface or Roche lobe, is shared between the two stars in a binary system. When one star expands beyond its Roche lobe, or the Roche surface shrinks into the star by loss of orbital angular momentum, any mass lying above the Roche surface can flow from one star to the other \citep{Pringle1985}.  
This process is called {\it Roche-lobe overflow}.  

The mass transfer naming convention is based on when, during its evolution, a star fills its Roche lobe.
\rgi{If mass transfer occurs during the core hydrogen burning phase of the donor, i.e. on the main sequence, it is called \textit{case~A} mass transfer. Alternatively, if mass transfer takes place while the donor is burning hydrogen in a shell, i.e.~on the first or red giant branch (RGB), it is called \textit{case~B}. After helium ignition, e.g.~on the asymptotic giant branch (AGB), mass transfer is referred to as \textit{case~C}.} 


Material that overflows can all be accreted by the companion, in which case the mass transfer is {\it conservative}. 
Alternatively, mass can be part accreted and part lost, a process called \textit{non-conservative} mass transfer. 
If the accretor gains enough mass, the mass ratio can invert, with the originally more massive primary becoming the least massive and dimmer star. 
Such systems are known as Algols, named after the prototype for the class ($\beta$~Persei, Table~1). 
\rgi{The post mass-transfer distribution of Algol binary properties, such as orbital periods, is an important test of mass transfer efficiency \citep[][Section~\ref{sssec:modelling-binary-interactions}]{vanResenbergen2011}.}

The evolutionary fate of \rgi{a} binary system depends crucially on whether the binary orbit shrinks or expands as a result of Roche-lobe overflow \citep{Soberman1997} and the response of the accreting star. When RGB or AGB stars expand and overflow their Roche lobe, they transfer mass to their companion ({\it cases B} and {\it C} mass transfer, respectively). If the accretor cannot accommodate the mass transferred to it, it too may fill its Roche lobe and a common envelope may form around the system (Section~\ref{sssec:modelling-binary-interactions-the-important-case-of-the-common-envelope-interaction}). A dynamical in-spiral phase follows, which typically results in a merger, or in the ejection of the common envelope and the emergence of a close binary, such as a sub-dwarf O or B binary\rgi{,} or the central star of a planetary nebula \citep[][Table~1; Section~\ref{ssec:the-origin-of-non-spherical-planetary-nebulae}]{Paczynski1976,Ivanova2013}. 

Mass transferred from one star to another carries angular momentum. Only about 10 per cent of a star's mass has to be accreted to spin the star to its break up rotation rate when it can accrete no more \citep{Packet1981}. A combination of tidal interaction and wind mass loss can prevent a star from spinning this fast and allow it to accrete \citep{deMink2014}\rgi{. The} excess angular momentum is \rgi{transferred} to the orbit \rgi{by tides} or \rgi{lost} in \rgi{a wind.} Tides also allow close systems, which undergo significant mass loss, to have rapidly spinning stars even until their death as supernovae and potential gamma-ray bursts \citep{Izzard2004a,Detmers2008}.

Magnetic fields are likely to alter the accretion pattern between two stars in at least some cases, particularly if the accretion is onto an evolved star, such as a white dwarf (WD) or a neutron star. The most well known and extreme example is perhaps offered by "polars" \citep[][Table~1]{Cropper1990}, where a highly magnetic WD (the record holder, AN~Ursae Majoris, has a field of $2.3\times 10^8$~G) is accreting from a main sequence companion. The accretion stream is channelled from the inner parts of the disk into the poles of the WD by the magnetic field or even directly onto the WD without the mediation of an accretion disk in the more extreme polars. Even more extreme are neutron stars accreting in high mass X-ray binaries that have magnetic fields in the range $10^{11-13}$~G \citep[e.g.,][]{Jaisawal2015}.

Finally, a new type of Roche-lobe overflow has been recently added to the nomenclature: {\it wind} Roche-lobe overflow. In these binaries the wind of the donor, for example a mass-losing AGB star, has a velocity within a small range of values from the value of the escape velocity at the Roche surface of the donor. In such cases even if the donor is not filling its Roche lobe, its wind naturally flows through the inner Lagrangian point and onto the accretor. This scenario was \rgi{investigated} by \citet{Mohamed2007} in order to explain the peculiar observations of the AGB star Mira that is clearly transferring mass to a companion despite an orbital separation so large that Mira cannot be possibly filling its Roche lobe \citep{Mohamed2012}.


\subsection{A list of binary classes}
\label{ssec:binary-classes}

Binary class nomenclature has arisen over the centuries in response to a range of binary discoveries achieved with a \rgi{variety} of observational techniques. Names have been coined for binaries discovered with a particular technique (e.g., radial velocity binaries), even when the systems detected \rgi{are} clearly a heterogeneous group from \rgi{an} evolutionary standpoint. As the wealth of binary interaction types started to emerge\rgi{,} classes that tried to divide the types of interactions emerged. According to the still oft-used \citet{Kopal1955} classification, binaries can be \rgix{either} \textit{detached} when neither star fills its Roche lobe, \textit{semi-detached} when one star fills its Roche lobe or in \textit{contact} when both stars fill their Roche lobe. Eventually\rgi{,} names emerged that collected objects thought to have evolved similarly (e.g., novae). This means that a given binary could be classified in more than one way (e.g., dwarf novae are semi-d!
 etached binaries).

In Table~\ref{tab:binary-classes} we present a list of binary class names that have been used in the literature, alongside their current, best interpretation. \rgi{W}e make no claim to completeness\rgi{, n}or do we dwell on each of these classes and on the subtleties of their interpretation. 
It is worth noting that a good classification system is one that is based purely on observed characteristics rather than interpretation. In Table~\ref{tab:binary-classes} Sequence~E stars are such an example. They appear on a specific locus of the period-luminosity diagram of Large Magellanic Cloud variables (Fig.~\ref{fig:sequenceE}), something that will remain the case, irrespective of the interpretation we give them. On the other hand, binaries are too diverse a group of objects to enforce strict classification rules. Some objects will be grouped by virtue of being interpreted as having a common evolutionary origin based on a whole range of disparate observations, even if that interpretation may turn out to be incorrect later on.


\begin{table*}

\newcolumntype{b}{X} 
\newcolumntype{m}{>{\hsize=.6\hsize}X}
\newcolumntype{s}{>{\hsize=.55\hsize}X}

\begin{tabularx}{\textwidth}{mbs}
\hline
\bf{Class name} & \bf{Interpretation/Comment} & \bf{Reference}
\\

\hline 



W Ursae Majoris & Low-mass (A to K type) contact binaries & \citealt{Terrell2012} \\

Cataclysmic~variables & Mass transferring WD+MS stars (novae, dwarf novae, polars [AM~Herculis and DQ~Herculis stars], etc.)&\citealt{Warner2003}\\
AM Canis Venaticorum & Helium-rich, mass transferring WD+MS stars & \citealt{Nelemans2005a} \\
Algols & The evolved star is less massive than the non-evolved star. Mass donor is stripped to reveal its nitrogen-rich core&\citealt{Budding2004}\\
Low mass X-ray binaries & NS or BH accreting from a low mass star& \citealt{Podsiadlowski2002}\\
High mass X-ray binaries & NS or BH accreting from a high mass star&\citealt{LiuQZ2006}\\
Sequence E stars  & RGB and AGB stars with close companion &\citealt{Nicholls2012}\\

Symbiotic binaries & WD accreting from an RGB or AGB giant &\citealt{Belczynski2000}, \citealt{Jorissen2002} \\
Wolf-Rayet + O binaries & Massive Wolf-Rayet + O star binaries & \citealt{Gosset2001}\\


Double degenerates & Often close, hard to understand, key to SN Ia & \citealt{Nelemans2005} \\
Blue stragglers & Brighter and bluer than cluster turnoff because of accretion&\citealt{Bailyn1995}\\
sdOB and extreme horizontal branch binaries & Stripped cores of RGB stars&\citealt{Drechsel2001} \\
Post-CE central stars of PN & CE interaction during the AGB & \citealt{Miszalski2009}\\ 
Post-AGB binaries & Post-AGB+MS binaries; P$\sim$100-1000~days with circumbinary disks & \citealt{VanWinckel2003} \\
FS Canis Majoris & Hot+cool close binaries exhibiting the B[e] phenomenon (?) & \citealt{Miroshnichenko2007}\\ 
Heartbeat stars & $\sim$1-2~\msun\ MS+MS binaries, short, eccentric orbits. Primary distorted at periastron & \citealt{Welsh2011} \\

\hline


\multicolumn{3}{l}{\it Chemically peculiar stars} \\
Barium stars & Single stars that accrete wind from a Ba-rich, AGB companion ($[\mathrm{Fe}/\mathrm{H}]\sim 0$) & \citealt{Merle2016} \\ 
 
CH/dwarf C stars & Single stars that accrete wind from a C-rich AGB companion ($[\mathrm{Fe}/\mathrm{H}]\sim -1$) & \citealt{Jorissen2016} \\

CEMP stars & As CH stars at lower metallicity ($[\mathrm{Fe}/\mathrm{H}]\lesssim -2$) & \citealt{Starkenburg2014} \\


\hline


\multicolumn{3}{l}{\it Outbursts} \\
Novae & Detonation on WD surface after accretion & \citealt{Bode2012} \\

Dwarf novae & State change of accretion disk around WD in WD+MS close binary  & \citealt{Osaki1996} \\

Symbiotic novae & Symbiotic binaries undergoing outbursts & \citealt{Mikolajewska2010} \\

Type Ia supernova & WD+WD merger or WD accreting from non-degenerate star & \citealt{Maoz2014} \\

Luminous blue variables & Massive stars, often binary e.g., $\eta$ Carinae, with outbursts, some maybe mergers & \citealt{Smith2011} \\

Gap transients & Outburst brightness between novae and supernovae & \citealt{Kasliwal2012} \\

Long gamma-ray bursts & Tidally locked binaries (?) & \citealt{Detmers2008} \\

Short gamma-ray bursts & NS+NS binary merger & \citealt{Berger2014} \\

\hline

\multicolumn{3}{l}{\it Single (merged), or apparently single binaries} \\


V~838 Mon, V1309 Scorpii & Stars observed in the process of merging (?) & \citealt{Tylenda2011b} \\

FK Comae stars & Rapidly rotating red giants, presumably merged & \citealt{Eggen1989} \\

R-type carbon stars & C-rich red giant, merged with WD & \citealt{Izzard2007} \\

R Coronae Borealis stars & Merger of a He and a CO WD & \citealt{Zhang2014}\\

Magnetic WDs & Dynamo induced by binary merger & \citealt{Briggs2015} \\
 

\hline

\multicolumn{3}{l}{Legend: MS = main sequence; WD = white dwarf; NS = neutron star; BH = black hole; CE = common envelope; } \\
\multicolumn{3}{l}{SN = supernova; sdOB = subdwarf O or B; CEMP = carbon enhanced metal poor.}\\
\end{tabularx}

\caption{Common names of classes of binaries and their likely interpretation (a "?" denotes an uncertain interpretation).}
\label{tab:binary-classes}
\end{table*}

\subsection{Binary stars that evolve in clusters}
\label{ssec:binary-stars-that-evolve-in-clusters}

Data on the binary populations of young clusters, such as those discussed in Section~\ref{ssec:binaries in cluster environments}, are fundamental for choosing realistic initial conditions for N-body simulations of clusters. These allow us to determine the evolution of the binary properties in those environments to understand older cluster observations and to probe evolutionary channels for a range of compact cluster binaries, including the formation of cluster specific binaries such as certain types of blue straggler stars (Section~\ref{ssec:binary-classes}). The binary population also  affects cluster dynamics and cluster observables in general, such as the width of the main sequence in colour-magnitude diagrams.

Data on the binary properties of the young cluster M~35 have been adopted by \citet{Geller2013} as input to their N-body model of the much older cluster N~188 (7~Gyr; \citealt{Meibom2009}), under the assumption that the binary characteristics of N~188 when it was young were the same as those of M~35. This study has revealed a wealth of information regarding the way in which cluster environment alters the binary population and, conversely, how the binary population alters the overall cluster properties.

The cluster binaries of N~188 are segregated in the centre of the cluster both in observations \citep{Geller2008} and simulations. The binary frequency was chosen to be 27 per cent ($P<10^4$~days) at the start of the simulation, in line with 24 per cent determined from observations of the young cluster M~35. This is extended in period to a total binary fraction of $\sim$60 per cent, which reduces to 53~per cent within the first 50~Myr of the simulated cluster life. By 7~Gyr, the binary frequency is 33.5$\pm$2.8 per cent, only slightly larger than at the start and in line with observations of N~188 ($29\pm3$ per cent; \citealt{Geller2012}). Twins, binaries with mass ratio close to unity, are not input into the simulations nor are they produced dynamically, pointing to the formation of twins being a process of star formation. The main sequence binaries largely maintain their characteristics throughout the 7~Gyr of cluster evolution (84 per cent of the solar type binaries do no!
 t change their characteristics), although wider binaries (period longer than $\sim$10$^{6.5}$~days) see their eccentricities increase over time. This may increase the number of binary interactions where the inner binary is perturbed by a tertiary in an eccentric orbit (Section~\ref{ssec:main-sequence-triples-and-higher-order-multiple-stars}). 

Finally, the number of blue straggler stars in NGC~188 is much larger than the number of blue stragglers predicted by the N-body models of \citet{Geller2013}. The models predict that almost all blue stragglers in the cluster derive from stellar collisions. The models also predict too many WD-main sequence binaries with circularised orbits, but with periods longer than the tidal circularisation period. It is therefore thought that a different criterion should be chosen in the models so as to have fewer common envelope interactions that form WD-main sequence binaries, and more stable mass transfer interactions that would generate more mass-transfer blue stragglers. This could be a solution to both  discrepancies between model and observations. Ultimately we need a better description of mass transfer to inform population synthesis models of which binaries enter a common envelope phase and which do not (Sections~\ref{ssec:types-of-binary-mass-transfer} and \ref{sssec:modelling-b!
 inary-interactions}).

\section{THE BINARY STAR TOOLKIT}
\label{sec:the-binary-star-toolkit}

A few observational tools have been\rgi{,} or are at present\rgi{,} particularly useful to binary star observers. The radial velocity technique exploiting high resolution spectrographs such as the {\it Fiber-fed Extended Range Optical Spectrograph} FAROS \citep[e.g.,][]{Setiawan2004} has been and will always be a tool of choice in studying binaries (and planets). Alongside this workhorse a host of new techniques and telescopes have been developed that present  new types of evidence (Section~\ref{ssec:the-toolkit-observationas-and-observatories}).

As for models, we distinguish modelling codes and techniques into two categories\rgix{,} which can function in unison. The first class numerically solves equations that describe the stars either hydrostatically or hydrodynamically. Individual binary systems are modelled over time and parameters such as the evolution of radius or luminosity, mass transfer rate, masses and kinematics of forming disks or jets are \rgi{output}. The second class includes population synthesis models. These can be more or less complex but \rgi{all} aim to model \rgix{the} population characteristics such as distributions of masses, luminosity functions or outburst rates. We \rgi{briefly} review some of these tools below (Section~\ref{ssec:the-toolkit-modelling-techniques-and-codes}).

\subsection{The toolkit: observations and observatories}
\label{ssec:the-toolkit-observationas-and-observatories}

\begin{figure}
\centering
\includegraphics[width=0.243\textwidth]{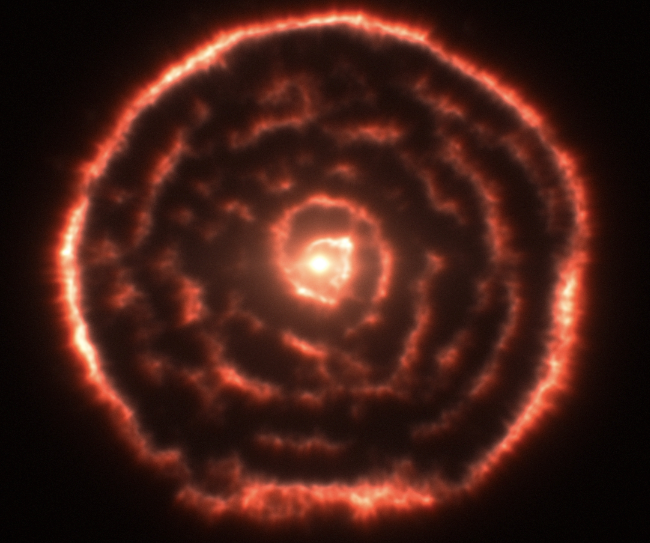}
\includegraphics[width=0.22\textwidth]{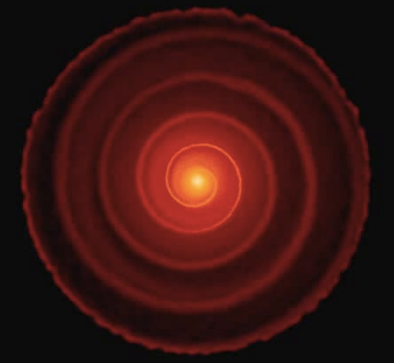}
\caption{Left panel: ALMA (Section~\ref{ssec:the-toolkit-observationas-and-observatories}) observation of the AGB giant R Sculptoris \citep{Maercker2012}. {\it Credit: ALMA Observatory}. Right panel: SPH hydrodynamic simulation of the system; the spiral wave requires a binary companion with a period of 445~yr sculpting the mass lost from the star. {\it Credit: Shazreen Mohamed, SAAO}. }
\label{fig:raquarii}
\end{figure}

The {\it Atacama Large Millimetre Array} (ALMA)\rgi{,} with \rgi{its} exquisite sensitivity and excellent spatial resolution\rgi{,} has \rgi{resolved} binary features such as spirals in giant star outflows\rgix{,} caused by orbiting companions \citep[][Fig.~\ref{fig:raquarii}]{Maercker2012}. It \rgi{also allows} us to characterise further, and lend corroborating evidence to the  phenomenon of wind Roche-lobe overflow  \rgi{(Section~\ref{sec:binary-classes}, also \citealt{Vlemmings2015})}. Large Keplerian disks around stars can be used to infer a binary past \citep[e.g.,][]{Bujarrabal2013} even when \rgi{a binary companion is not seen}. Finally, the detection of magnetic fields strengths and geometries in giant stars \citep[e.g.,][]{PerezSanchez2013} is a fundamental step to understand how they are generated and the interplay between duplicity and magnetic fields \citep{Nordhaus2006,Nordhaus2007}.

Optical and near infrared interferometry has been particularly successful in the study of binaries. The {\it Center for High Angular Resolution Astronomy} (CHARA) Array, for example, resolved the orbit of double-lined spectroscopic binary 12~Persei \citep{Bagnuolo2006}, allowing precise mass estimates that can later be used as calibrators of other systems. It was also used to resolve the inner orbit in hierarchical triples, including Algol systems \citep{OBrian2011,Baron2012}, and even succeeded in resolving massive Wolf-Rayet type binaries \citep{Richardson2016}.

Interferometry carried out with the {\it Very Large Telescope Interferometer} (VLTI) was the one to resolve disks around post-AGB stars \rgix{for the first time} \citep[][Section~\ref{sssec:modelling-binary-interactions-the-important-case-of-the-common-envelope-interaction}]{Deroo2006}, not to mention a host of disks and tori nested inside the cores of pre-planetary nebulae \citep[e.g.,][]{Chesneau2007} or in newly exploded stars thought to be the product of a merger \citep[e.g.,][Section~\ref{sec:transients}]{Chesneau2014}. Today, thanks to \rgix{the new-generation instrument} {\it Spectro-Polarimetric High-contrast Exoplanet REsearch}, SPHERE, detailed observations are being taken of AGB systems such as LB~Pup where a (presumed) binary \rgi{causes} a bipolar outflow \citep{Kervella2015}.

Also of note is the {\it Kepler Space Telescope} that stared at a small patch of sky reaching micro-magnitude variability detections. {\it Kepler} has had a tremendous impact on the characterisation of binaries. A large new catalogue of assorted eclipsing binaries was compiled \citep{Prsa2011},  subtle, rare or previously unseen phenomena, like the heartbeat stars were discovered (Section~\ref{ssec:heartbeat-stars}), rotation rates were measured in post-common envelope WDs \citep{Hermes2015} or in single WDs rotating so fast that they must be the product of mergers \citep{Handler2013}, the phenomenon of Doppler beaming was detected in close compact binary systems including some double degenerates \citep[e.g.,][]{vanKerkwijk2010,DeMarco2015}.  Finally, speckle interferometry has been useful to map a variety of binary stars, including the dusty environments of Wolf-Rayet-O star binaries (Section~\ref{sssec:wolf-rayet-stars}).

\rgi{Surveys capable of observing short-timescale, transient sources in great detail form a pillar of 21st century astronomy and directly affect binary-star observations. These surveys observe in wavelengths from radio to gamma rays and hence probe objects in exquisite detail. They serve as early-time alert mechanisms for deep, multi-wavelength follow-up observations. Many binary-star phenomena are directly accessible to these surveys, such as gamma-ray bursts, novae, stellar mergers, tidal disruptions, supernovae and, of course, gravitational waves, as will be discussed in Section~\ref{sec:transients}.}

\subsection{The toolkit: modelling techniques and codes}
\label{ssec:the-toolkit-modelling-techniques-and-codes}

\begin{table*}

\begin{tabular}{lll}
\hline 
\textbf{Name} & \textbf{Reference} & \textbf{Special features}\tabularnewline
\hline 

\multicolumn{3}{l}{\it Detailed binary stellar evolution codes} \\

{\sc stars}$^{a}$ & \citealt{Stancliffe2009} & Semi-Lagrangian mesh, nucleosynthesis and structure  \\
&&solved simultaneously, Eggleton's tides\tabularnewline
{\sc twin}$^{a}$ & \citealt{Glebbeek2008}   & Same as above \tabularnewline
{\sc bs}$^{a}$ & \citealt{Stancliffe2009} & Same as above \tabularnewline
{\sc rose$^{a}$} & \citealt{Potter2012}   & Rotation, magnetic fields\tabularnewline
{\sc mesa} & Paxton et al. 2015& Open source, community driven \tabularnewline
{\sc bec$^{b}$} & \citealt{Yoon2010}  & Rotation, magnetic fields, tides\tabularnewline
{\sc binstar}$^{b}$ & \citealt{Siess2013} &   s-process, eccentric orbits, tides\tabularnewline
 --                   $^{b}$ & \citealt{Podsiadlowski2010} &   X-ray binaries, common envelopes, mergers\tabularnewline
{\sc pns}$^{c}$ & \citealt{DeDonder2004}  & Stellar population grids\tabularnewline
 --                        & \citealt{Benvenuto2004} &   Simultaneous solver, nucleosynthesis \\
 &&Formation of helium WDs\tabularnewline

\hline  
\multicolumn{3}{l}{\it Synthetic binary stellar evolution codes}\\
 
{\sc bse}$^{d}$ & \citealt{Hurley2002} & Open clusters (in N-BODY6)\tabularnewline
{\sc binary\_c}$^{d}$ & \citealt{Izzard2009} & Nucleosynthesis and TPAGB, API, community driven\tabularnewline
{\sc startrack}$^{d}$ & \citealt{Belczynski2007} &  Massive binaries, black holes; low mass binaries, \\
&&type Ia supernovae\tabularnewline
{\sc BiSEPS}$^{d}$ &  \citealt{Willems2002,Willems2004}&\tabularnewline
{\sc SeBa}$^{e}$ & \citealt{Toonen2013} &  SNeIa, common envelope evolution\tabularnewline 
{\sc IBiS} & \citealt{Tutukov1996} & \tabularnewline
{\sc scenario machine} & \citealt{Lipunov2009}   & Massive binaries\tabularnewline

\hline 
\multicolumn{3}{l}{\it Hybrid binary stellar evolution codes and hydrodynamics}\tabularnewline

 --                   & \citealt{Chen2014}   & {\sc bse} + {\sc mesa} hybrid code\tabularnewline
{\sc amuse} &  \citealt{PortegiesZwart2009}  & {\sc twin} + 3D SPH$^g$ ({\sc fi}; \citealt{Pelupessy2012}) + \\
&&N-body ({\sc huano}; \citealt{Pelupessy2005}) \tabularnewline

\hline
\multicolumn{3}{l}{\it Hydrodynamic codes}\tabularnewline

{\sc djehuty} & Bazan et al. 2003&  Arbitrary Lagrangian-Eulerian code used for stars in 3D \tabularnewline
{\sc co$^5$bold} & Freytag et al. 2002&  Grid code used for stars in 3D \tabularnewline
- & \citealt{Woodward2003}&  Grid code used for stars in 3D \citep{Herwig2014}\tabularnewline
{\sc flash} &\citealt{Fryxell2000}& Grid AMR$^f$ code adapted for CE interactions by \\
&&\citet{Ricker2008}\tabularnewline
{\sc enzo} &  \citealt{Bryan2014}  &  Grid AMR$^f$ code adapted for CE interactions \\
&&by Passy et al. (2012)\tabularnewline
{\sc snsph} &  \citealt{Fryer2006}  &  SPH$^g$ code adapted for CE interactions \\
&&by Passy et al. (2012)\tabularnewline
{\sc --} &  \citealt{Lajoie2011}  &  SPH$^g$ code based on \citet{Bate1995}, adapted for eccentric \\
&&mass transfer interactions \\
{\sc starsmasher} & \citealt{Lombardi2011}   &  SPH$^g$ code adapted for CE interactions \\
&&(e.g., Nandez et al. 2016) \tabularnewline
{\sc phantom} & \citealt{Lodato2010}   &  SPH$^g$ code adapted for CE interactions \\
&&by Iaconi et al. (2016) \tabularnewline
{\sc arepo} & \citealt{Springel2010}   &  Moving mesh code adapted for CE interactions \\
&&(e.g., Ohlmann et al. 2016) \tabularnewline
{\sc gadget} & \citealt{Springel2005}   &  SPH$^g$ code adapted for binary interactions\\
&& by  \citet{Mohamed2012b} \tabularnewline
{\sc mpi-amrvac} & \citealt{Porth2014}   &  Grid AMR$^f$ code for wind-wind interactions \\
&&\citep[e.g.,][]{Hendrix2016} \tabularnewline
\hline
\multicolumn{3}{l}{Code family: $a$ \citealt{Eggleton1971}, $b$ \citealt{Kippenhahn1967}, $c$ Paczynski , $d$ {\sc bse/sse}, $e$ {\sc SeBa/sse}. } \\
\multicolumn{3}{l}{$f$ AMR = adaptive mesh refinement; $g$ SPH = smooth particle hydrodynamics. } \\
\end{tabular}
\caption{A list of some of the major computational programs used in the study of binary evolution.}
\end{table*}

\subsubsection{Modelling binary interactions}
\label{sssec:modelling-binary-interactions}

Modelling \rgi{a} single star is a complex task, despite the fact that, by and large, the evolution of a single star is determined only by its mass\rgi{,} composition and rotation rate. With some reasonably well-justified simplifications, such as the assumption of spherical symmetry, hydrostatic equilibrium and the mixing length theory for convection, stellar evolution is tractable on reasonable timescales and a range of 1D stellar evolutionary codes exists. Ideally binary interactions should be modelled in 3D, where both stars are modelled  with the same accuracy as in 1D and where the interaction is tracked by solving the Euler equation using self-gravity, full radiation transport and magnetic fields. Such complexity is at the moment beyond the realm of possibility, because of the vast range of time and size-scales that needs to be resolved. Some of the codes and code families have been listed in Table~2.

{\it Single star models using 3D hydrodynamic codes}. Parts of (single) stars can be modelled in 3D, for example, to model convective and rotational mixing \citep{Meakin2007,Cristini2015}. Full 3D hydrodynamical models of stars have been constructed with the {\sc djehuty} code at Livermore \citep{Bazan2003}, but they are extremely computationally intensive. 2D, hydrostatic stellar evolution is also starting to be explored as a natural stepping stone to full 3D modelling \citep{EspinosaLara2013}. Convection in giant stars was studied using 3D models by \citet{Meakin2007} and more recently by  \citet{Chiavassa2011} with {\sc co$^5$bold} \citep{Freytag2002} at relatively low resolution, but high enough for a meaningful comparison with VLTI observations of \citet{Wittkowski2016}. This revealed the size of the modelled convection plumes to be approximately correct. \citet{Herwig2014} modelled hydrogen entrainment in giant stars in 3D revealing the need for 3 dimensions to model A!
 GB thermal pulse nucleosynthesis.

{\it Binary interaction models using 1D implicit codes}. One-dimensional stellar evolution codes are used to model binary interactions, with binary phenomena such as accretion accounted for after parameters such as the orbital separation and accretion rates are calculated analytically, or guided by separate simulations with 3D codes (see below). 
Many such codes exist, of which there are a few families. The Eggleton codes are based on the single-star code of \citet{Eggleton1972}. Unique features include a non-Lagrangian moving mesh, which reduces the computational time involved in converging a stellar model, with the inclusion of some unwanted numerical diffusion. Modern versions of this code include {\sc twin} \citep{Glebbeek2008}, {\sc stars} and {\sc bs}  \citep{Stancliffe2008}. All include mass transfer and tidal interactions, with the most modern version of {\sc bs} also including magnetic field generation \citep{Potter2012}.

Another commonly used binary-star code is based on the original Kippenhahn code, \rgi{exemplified by} the {\it Bonn Evolutionary Code} ({\sc bec}, \citealt{Heger2000,Yoon2010}). This includes parametrised rotational mixing, magnetic fields, mass transfer and tidal interactions. The {\sc binstar} code of the (French-speaking) Brussels group \citep{Siess2013} also derives from this original code base, although it has been updated to include, for example, the physics of mass transfer in eccentric systems \citep{Davis2013}. The Flemish-speaking Brussels group also has a binary-star code, called the {\it Population number synthesis} ({\sc pns}; \citealt{DeDonder2004}), which it uses for both detailed evolution and population synthesis (Section~\ref{sssec:modelling-binary-populations}). 

The newest addition to the selection of binary-star codes is that of the {\sc mesa} group \citep{Paxton2015}. This combines the widely-used {\sc mesa} single-star code with binary-star physics. Among its advantages, {\sc mesa} was designed from the beginning by a software engineer, so it is relatively easy to use and develop.  

{\it Binary interaction models using 3D hydrodynamic codes.} Hydrodynamic models of binary interactions do exist, but they must make a number of \rgi{simplifying} assumptions and be guided by analytical considerations, in particular if they include self gravity of the gas, something that will make simulations much slower. They represent  the stars as simple hot spheres of gas in hydrostatic equilibrium. Example simulations  are those of \citet{Lajoie2011b} who modelled eccentric interactions between main sequence stars, deriving parameters such as the time of maximum mass transfer rate compared to the time of periastron passage. These models can be compared with analytical models of this phase such as those of \citet{Sepinsky2009}. The WD-WD merger simulations of \citet{Staff2012} carried out to simulate the formation of R~Coronae~Borealis stars from which merger temperatures and timescales can be interfaced with 1D stellar structure models to determine the nucleosynthetic s!
 ignature of these mergers \citep{Menon2013}.  

A different class of binary interactions can be modelled without using self gravity of the gas, but where a gravitational field is imposed, such as that produced by an object embedded in the gas. Wind-wind collision models have been performed over the decades to understand all kind of phenomena associated with single stars (e.g., in planetary nebulae; \citealt{GarciaSegura1999}). Similar techniques can be adopted in the study of wind-wind collisions in binary systems such as for example the collision between the wind of a Wolf-Rayet star and that of an O star companion, as we describe in detail in Section~\ref{sssec:wolf-rayet-stars}.  \citet{Hendrix2016} listed the history of such models starting with the models of \citet{Stevens1992} and ending with those of \citet{BoshRamon2015}. They also presented a model of the pinwheel nebula WR98a carried out with the code {\sc mpi-amrvac} \citep{Porth2014}. These models aim to study the wind interaction region as accurately as possi!
 ble to understand how dust forms. Pinwheel nebulae are chief dust producers, despite the relatively hostile environments and understanding these interactions contributes to the larger understanding of the dust budget of the Universe. However, these simulations are not aimed at understanding the evolution of the binary per se, although without doubt such systems will have quite an interesting life as both the stars are due to explode as core collapse supernovae at some point.

Somewhat similarly, the simulations of \citet{Booth2016} using the SPH code {\sc gadget} \citep{Springel2005} in the adaptation of \citet{Mohamed2012b} were used to study the circumstellar environments of symbiotic novae (WDs accreting from giant stars' winds; Table~1). Such systems can in principle be progenitors of type Ia supernovae and their circumstellar environment could cause observed absorption line variability first observed in supernova type Ia 2006X \citep{Patat2007}. The same code was used to simulate spiral shocks imprinted by a wide binary companion in a long orbit with an AGB star, as seen by ALMA in Sec.~\ref{ssec:the-toolkit-observationas-and-observatories}, see also Fig.~\ref{fig:raquarii} (for a similar approach see also the work of \citet{Kim2012}).

A  creative technique is that of using a range of different codes as well as analytical approximations in unison. An example is the work of \citet{deVries2014}, who modelled a tertiary star in a triple system overflowing its Roche lobe and transferring mass to the compact binary in orbit around it. To do so, they used 1D stellar structure codes, a 3D hydrodynamics code and an N-body integrator handled via the  {\it Astrophysical Multipurpose Software Environment} \citep[{\sc amuse};][]{PortegiesZwart2009}.

\subsubsection{3D hydrodynamic models of the important common envelope binary interaction}
\label{sssec:modelling-binary-interactions-the-important-case-of-the-common-envelope-interaction}
\begin{figure*}
\centering
\includegraphics[width=0.75\textwidth]{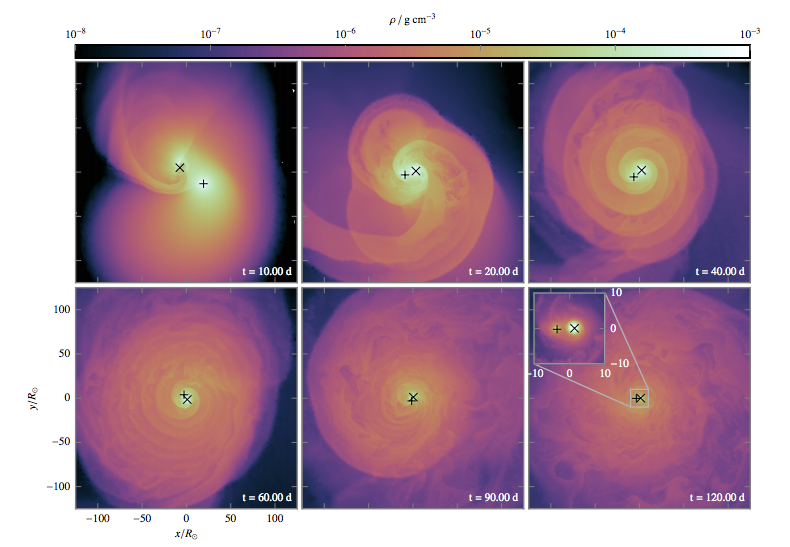}
\caption{A series of density slices at six different times along the orbital plane during a 3D, hydrodynamic simulations of a common envelope in-spiral (Section~\ref{sssec:modelling-binary-interactions-the-important-case-of-the-common-envelope-interaction}) of a 1-\msun\ companion in the envelope of a 2-\msun\ RGB star. The X marks the position of the companion, the plus symbol marks the position of the RGB star's core. The insert  shows a central region of approximately 20~\rsun. The colour scale ranges between $10^{-6}$ and $10^{-3}$~g~cm$^{-3}$. {\it Credit: image adapted from figure 3 of \citet{Ohlmann2016}.}}
\label{fig:ohlmann2016}
\end{figure*}

Common envelope (CE) interactions deserve a special mention. The idea of the CE interaction was put forth by \citet[][who credits other authors for the original idea, such as \citet{Webbink1975} and a private communication by J. Ostriker, among others]{Paczynski1976} to explain the binary V~471~Tau a pre-cataclysmic variable with an orbital separation much smaller than the presumed radius of the progenitor of the WD primary. Many classes of objects are in a similar situation, including cataclysmic variables, low and high-mass X-ray binaries and the progenitor of many classes of stellar mergers such as type Ia supernovae, neutron star and black hole mergers. For a recent review on the CE interaction see \citet{Ivanova2013}. See also \citet{Iben1993}, \citet{Livio1988} and \citet{Taam2000}.

Our understanding of the interaction is partial and at the moment we cannot predict the relationship between pre-CE and post-CE populations. There have been many papers that have emphasised the issues arising from this problem such as that of \citet{Dominik2012} who analysed the impact of the uncertainties on the CE phase on the predicted merger rates of WDs, neutron stars and black holes, or the work of \citet{Toonen2013} who analysed the impact of different CE prescriptions on the characteristics of post-CE binaries in general.

One of the main issues is our ignorance of the efficiency of the energy transfer between the orbit and the envelope of the primary. In fact this problem is even more complicated by realising that the orbital energy is not the only source of energy potentially available and other sources, such as recombination energy, can be unlocked by the interaction. Ultimately this efficiency parameter has been used as a single number, sometime alongside a second parameter that changes depending on the specific structure of the primary. Sometime a second efficiency factor is used in combination with sources of energy other than orbital energy \citep{Han1995}. Studies aiming at finding what the efficiency of the CE might be have used known post-CE systems for which the pre-CE configuration could be reconstructed \citep[e.g.,][]{Zorotovic2010,DeMarco2011} or have used population synthesis codes with different CE efficiency prescriptions in the hope of using population constraints to constra!
 in the efficiency  \citep[e.g.,][]{Politano2007}.

Another technique to study the CE phase is 3D hydrodynamic simulations and early work includes the simulations of \citet{Yorke1995}, \citet{Terman1996}, \citet{Sandquist1998} and \citet{Sandquist2000}. The dynamical phase of the in-spiral takes place over a short dynamical timescale (hundreds of days). The envelope is lifted away from the binary and the orbital distance stabilises. However, only a small fraction of the envelope is typically ejected. The CE efficiency parameter values calculated by these simulations are not quite the same as those needed by population synthesis because the envelope is not ejected and because there cane be no inefficiency due to radiation as the codes are adiabatic.

The number of models of the common envelope phase has increased in the recent years, but they are still far from being predictive \citep[e.g.,][]{Ricker2008,Passy2012,Ricker2012,Ohlmann2016,Staff2016,Ohlmann2016b,Iaconi2016}. The main issue is that the dynamical in-spiral phase is unable to eject the envelope in most models, which leaves the question of whether \rgi{evolution after the dynamical in-spiral phase} holds the key to the final configuration of the binary \citep{Ivanova2013}. The phase following the fast in-spiral takes place over a longer, thermal timescale and \rgi{is} very difficult to model within the same  simulation that models the faster, dynamical in-spiral \citep{Kuruwita2016,Ivanova2016}. 

Recently, the inclusion of recombination energy in the energetics of the dynamical interaction has enabled a set of simulations to unbind the entire envelope \citep{Nandez2015,Nandez2016}. The main question is how much of this energy is available to eject the envelope instead of being radiated away. The main argument for the energy availability is that the hydrogen and even more so the helium recombination fronts form deep within the star when the envelope starts expanding and cooling. Even if the optical depth decreases in front of the recombination front, it is argued that it is unlikely that all of the recombination energy would leak out. This seems a valid argument, but an actual test of the fraction of recombination energy that leaks away will have to await a full radiation transport treatment in the codes.

It has also been suggested that the formation of jets takes place during the CE \citep[][]{Soker2004b,Nordhaus2006} and that this may aid in ejecting the envelope. It is likely that this can happen, but it is not obvious that it would happen under all circumstances. Even the relatively homogeneous group of post-CE planetary nebulae displays jets only in a minority of cases. When we do see jets in post-CE planetary nebulae, their kinematics can be compared with the kinematics of the bulk of the planetary nebula, which is assumed to be the ejected CE, from which we deduce that jets can be ejected immediately preceding or immediately following the CE ejection \citep{Tocknell2014}. Given the uncertainties one could argue that jets could be launched also during the dynamical in-spiral in some cases. However, given the current uncertainties on the theory of jet launching (Section~\ref{ssec:jets}) it is far from clear when and how these jets would form. Were they to form, however, !
 it is likely that they would play a major role in the dynamics and energetics of the CE ejection \citep{Soker2004}.

Although many uncertainties surround the CE phase, we assume that a fast phase of dynamical in-spiral does indeed take place, possibly preceded and followed by much longer phases. It is likely that a tidal phase takes place before the in-spiral leading up to the moment of Roche lobe contact and it is also likely that a post-in-fall phase follows, on a longer thermal timescale, regulated by thermal adjustments of the star(s) as well as possibly by some envelope infall \citep{Kuruwita2016}. 

There is, however, a class of binaries that contradicts the belief that a fast in-spiral {\it always} leaves behind a close binary or a merger. Some post-AGB stars have main sequence companions in orbits with periods between $\sim$100 and $\sim$2000 days \citep[see Table~1][]{VanWinckel2003,VanWinckel2009}. The binaries with the shortest periods have circular orbits, while the longer period binaries can have quite eccentric orbits. The shortest period binaries must have gone through a CE phase, which did not lead to a dramatic in-spiral. Suggestions such as the "grazing CE" idea of \citet{Soker2015} rely on a series of mechanisms working in unison, such as high accretion rates onto the companion, accompanied by a jet production that can remove mass and energy early on. It remains to be seen whether they can operate in these cases. For the time being, we know that these objects have circumbinary tori but usually no visible nebula with the exception of one system, the Red Rect!
 angle \citep{VanWinckel2014}.

The ultimate goal of CE simulations is to predict the parameters of post-CE binaries and mergers as a function of pre-CE binary parameters. Such predictions can be then parametrised for the use of population synthesis codes, which interpret the bulk characteristics of entire populations (Section~\ref{sssec:modelling-binary-populations}). An attempt at such parameterisation was carried out by \citep{Ivanova2016}, but their computational efforts need further verification steps before they can be generally adopted.

\subsubsection{Modelling binary populations}
\label{sssec:modelling-binary-populations}

The binary-star parameter space is much larger than that of single stars. This has led to binary star modelling taking two directions. Either a full, detailed binary stellar evolution code is used to model few stars (Section~\ref{sssec:modelling-binary-interactions}), or a simplified synthetic code covers a larger parameter space with more stars. Both techniques are called population synthesis, which should not be confused \rgi{with the related field of} population synthesis of integrated spectra from unresolved stellar populations.

The detailed model approach uses the binary codes described earlier (Section~\ref{sssec:modelling-binary-interactions}). The Brussels code {\sc pns}, for example, performs population syntheses by interpolating on a grid of detailed binary-star models \citep{deDonder1997}. Internal stellar structure is thus known, and the models are those of true binary stars. The code used by, e.g., \citet{Han2002}, also interpolates on a grid of pre-calculated detailed models \citep{Han1995}. The {\sc bpass} model set (\citealt{Eldridge2008}, \citealt{Stanway2015}), calculated with the {\sc stars} code, has been used to model many aspects of binary stars such as individual binary stars, supernova progenitors and spectra of high redshift galaxies.

The \rgi{synthetic} approach is much faster but less accurate. There are a few rapid synthetic binary population codes: the most prominent are {\sc bse} and {\sc SeBa}. Both are based on the {\it Single-Star Evolution} code \citep[{\sc sse};][]{Hurley2000}, which is based on detailed single-star models \citep{Pols1998}. Fitting functions approximate the stellar radius, luminosity, core mass and other parameters as a function of time. Binary star evolution is added to include mass transfer, common envelope evolution, tidal interactions, magnetic braking and stripped objects such as helium stars and WD.

The {\sc bse} code is available for download and it is embedded in {\sc nbody6} \citep{Aarseth2003}. The {\sc binary\_c} code \citep{Izzard2004a,Izzard2006,Izzard2009}, adds nucleosynthesis, updated physics, a suite of software for population synthesis and visualisation, and an \rgi{API (Application Programming Interface)}. The {\sc startrack} code is based on the {\sc bse} \rgi{algorithm}, with \rgix{an} emphasis on massive stellar evolution and X-ray binaries \citep{Belczynski2007}, although it is also used for intermediate mass stars, particularly in the field of type Ia supernovae \citep{Ruiter2014}. The {\sc SeBa} code is based on {\sc sse}, but implements Roche-lobe overflow with an algorithm based on radius exponents $\zeta=\partial{\,\ln M}/\partial{\,\ln R}$ \citep{PortegiesZwart1996}.

Code verification and validation are key to justifying the use of simplified models in place of more detailed and computationally expensive codes. The POPCORN project to investigate \rgi{type Ia supernova} progenitors \citep{Toonen2014} compares the {\sc binary\_c}, {\sc SeBa}, {\sc startrack} and the Brussels codes. The choice of input physics is the main difference between results from the codes. Comparison between {\sc binary\_c} and {\sc bec} led to an improved model for Roche lobe overflow in {\sc binary\_c} \citep{Schneider2014}. Massive-binary mass transfer in the two codes now agrees quite well, while the original {\sc bse} formalism predicts often quite different final masses and evolutionary outcomes.

Finding non-ambiguous ways to compare population synthesis models to observations is a key step for successful validation. A great example is the comparison between the modelled numbers and the observations of blue stragglers and WD-main sequence circularised binaries discussed in Section~\ref{ssec:binary-stars-that-evolve-in-clusters}. 

The disadvantage of synthetic modelling is that single stellar evolution tracks only approximate real binary stars. To solve many problems, this is good enough. There are, however, occasions when the lack of a true binary-star model is problematic, e.g.,~when accreted mass significantly changes the composition or size of a star. That said, a factor \rgi{of about} $10^7$ gain in speed allows a huge parameter space to be explored with a \rgi{synthetic} model even though care must be taken when interpreting the results and estimating systematic errors. Such parameter spaces are too large for detailed stellar evolution at present, but this will change in the future. Hybrid approaches are a pragmatic \rgi{step forward} \citep{Nelson2012,Chen2014}.

\section{BINARIES AS LABORATORIES}
\label{sec:binaries-as-laboratories}

Physical phenomena \rgi{caused by a companion star are} better understood as additional binary parameters are measured with increasing accuracy. Here we \rgi{comment} on two aspects of astrophysics, which are of broad interest and applicability and which can be best studied in binary systems. The first is disks and jets, the second is stellar structure.

\subsection{Disks and jets}
\label{ssec:jets}
\begin{figure*}
\centering
\includegraphics[width=0.6\textwidth]{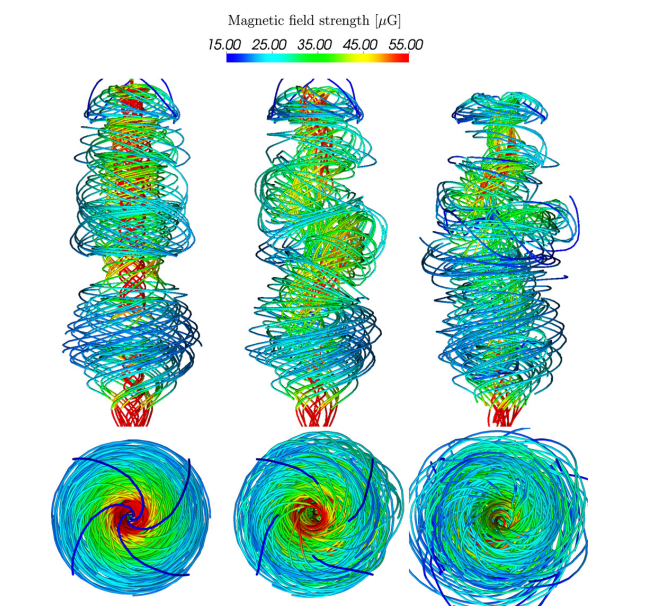}
\caption{Central magnetic field lines in three simulations of magnetic "tower" jets, a distinct type of jet to the classical magneto-centrifugally launched jet of \citet{Blandford1982}. Jet launching mechanisms are widely applicable to a range of astrophysical environments and can be studied observationally using interacting binary stars (Section~\ref{ssec:jets}). The three jets are calculated under different assumptions (left: adiabatic, centre: the rotating, right: cooling magnetic towers). The bottom panels show an upper view, pole-on. Open field lines are a visualisation effect. {\it Credit: image adapted from figure 6 of \citet{HuarteEspinosa2012}.}}
\label{fig:magnetictower-fig6}
\end{figure*}

The importance of jets is not limited to binary interactions. 
They are also important in star formation, where the jet is driven by a disk of material accreting from the interstellar medium\rgi{. Jets also} regulate galactic engines, where they are observed in active galactic nuclei.  Although the formation and launching of jets remains the subject of debate, the most commonly used launching model is that of \citet{Blandford1982}. An accretion disk is threaded by a magnetic field and as mass loses angular momentum and moves from the outer to the inner disk, a fraction is shot out and collimated in a direction approximately perpendicular to the plane of the disk. 
The nature of the viscosity that allows gas to accrete is not clear, but it might be provided at least in part by the very same magnetic field that \rgi{is} responsible for the jet collimation \citep{Wardle2007}.

Measured jet parameters\rgi{,} such as energies and momenta, help constrain the engine that launches the jet. Kinematic measurements of the highly collimated molecular outflows typical of pre-planetary nebulae show that radiation cannot be responsible for accelerating, nor collimating the gas, because the measured linear momenta exceeds by 2-3 orders of magnitudes what can be driven by radiation \citep{Bujarrabal2001}. On the other hand jets from accretion disks formed during binary activity could explain the observations \citep{Blackman2013}. Other interesting cases are the collimated structures seen in planetary nebulae with post-common envelope central star binaries studied by \citet[][Section~\ref{ssec:the-origin-of-non-spherical-planetary-nebulae}]{Jones2014,Jones2014b}. Their kinematics were used by \citet{Tocknell2014} to impose constraints on the common envelope interaction energies, timescales and magnetic fields.  

It is possible that the mechanism that ejects mass in some binaries is different from the jet launching mechanism of \citet{Blandford1982}. Magnetic pressure-dominated jets from tightly wound fields (magnetic "springs" or "towers"; \citealt{LyndenBell2003,LyndenBell2006}) can arise under typical conditions, and this could result in different outflow powers and observable characteristics of the asymptotically propagating jet. Magneto-centrifugal jets are magnetically dominated only at the base, and gas accelerated from the disk eventually dominates the magnetic energy at large outflow distances. In contrast, the magnetic "spring" jets can in principle be magnetically dominated \rgi{out} to much larger scales \citep[][Fig.~\ref{fig:magnetictower-fig6}]{HuarteEspinosa2012}.

Disks, jets and outflows from binaries can be studied in great detail and possibly even be observed as they form in transients (Section~\ref{sec:transients})\rgi{. Such observations} will soon allow us to put together a more satisfactory picture of their origin. 

\subsection{Stellar and tidal parameters from \rgi{asteroseismology} of heartbeat stars}
\label{ssec:heartbeat-stars}

\begin{figure*}
\centering
\includegraphics[width=0.7\textwidth]{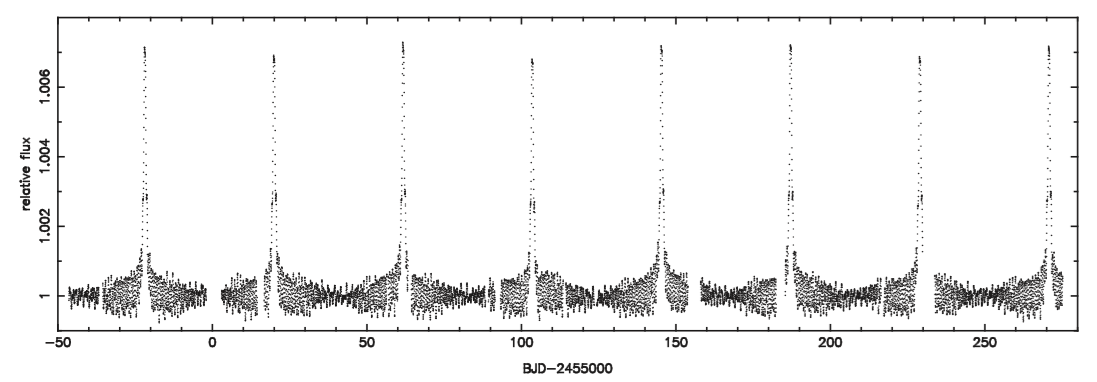}
\caption{The detrended and normalised {\it Kepler Space Telescope} light curve of heartbeat binary KOI-54. Heartbeat stars are giants with companions in eccentric orbits (Section~\ref{ssec:heartbeat-stars}). The companion "plucks" the giant at periastron passage and the giant "rings", producing a distinctive spike pattern that is used to study a range of physical properties from tidal dissipation to giant envelope structure. {\it Credit: image adapted from figure 1 of \citet{Welsh2011}}.}
\label{fig:Walsh+11-heartbeat}
\end{figure*}

Heartbeat stars are \rgi{low- and intermediate-mass} main sequence stars \citep{Smullen2015} and giants \citep{Hambleton2013} with nearby main sequence companions in eccentric orbits. At periastron the stars exert a tidal force on each other that distorts their envelopes and induces oscillations that are revealed in \rgi{their} lightcurve\rgi{s}\footnote{The characteristic look of the lightcuve with one strong pulse followed by smaller, ringing pulses looks like an electrocardiogram of a beating heart (Fig.~\ref{fig:Walsh+11-heartbeat}).} \citep[][Fig.~\ref{fig:Walsh+11-heartbeat}]{Welsh2011}. \rgi{About 130 were discovered} thanks to the high precision photometric observations of {\it Kepler} \rgi{\citep{Hambleton2013}}. 
Such stars have been used to constrain further stellar parameters, because the induced pulsations allow us to use \rgi{asteroseismological} techniques. 

The interplay \rgi{between} natural stellar pulsations and the periodic plucking action of an eccentric companion \rgi{complicates} the analysis of some \rgi{binary} systems  \citep[e.g.,][]{Gaulme2013}\rgi{.} 
However, with the increased availability of high precision variability observations, well constrained complex models will be possible, and these stars will become useful probes of stellar parameters. 

Another interesting application of the heartbeat stars is the possibility of measuring the tidal dissipation parameter $Q$ \citep{Goldreich1963}\rgi{. 
The} relationship between the \rgi{lightcurves} of heartbeat stars and their radial velocities can\rgi{,} in principle\rgi{,} constrain the angle between \rgi{tidal} bulges and the line connecting the two centres of mass \rgi{of the stars} \citep{Welsh2011}. This is an important and uncertain parameter in models of tides in stars (Section~\ref{sssec:modelling-binary-interactions}).

\subsection{Tests of extreme physics}
\label{ssec:tests-of-extreme-physics}

Close binaries containing neutron stars are another laboratory provided by binary-star evolution in which matter is at extreme temperatures and densities currently irreproducible on Earth. These binaries contain two compact, degenerate stars, usually a neutron star and a WD. In the double pulsar PSR J0737-3039 both stars are neutron stars. Timing of pulsar radio emission allows extremely precise measurement of the theory of General Relativity \citep{Kramer2006}. The properties of the neutron stars in binaries, such as masses and radii, constrain the unknown neutron star equation of state. When neutron stars merge, they not only make r-process elements \citep{Rosswog2014,Shen2015}, but also gravitational waves that may be detectable in the near future \citep[][Section~\ref{ssec:gravitational-wave-sources}]{Agathos2015}.

\subsection{Wind-polluted stars and ancient nucleosynthesis}
\label{ssec:wind-polluted-stars}

For some time, a major problem in stellar physics was that barium stars, red giant stars with atmospheres enriched in barium, were too dim ($L \sim 100-1000~\mathrm{L}_\odot$) to have manufactured the observed barium, something that should happen at luminosities higher than approximately  $10^4~\mathrm{L}_\odot$.  It was later discovered that the barium stars are binaries polluted by a companion that \rgi{had previously manufactured the} barium. The companion became a WD, which is today too dim to see \citep{Merle2016}.
This solved the mystery (\citealt{McClure1980}; \citealt{McClure1983}).

In barium stars the companion accreted the barium not \rgi{by} Roche lobe overflow, but from the barium-rich wind of the AGB star. This wind is gravitationally focused by the companion, often leading to significant accretion \citep{Edgar2004}. If the accreting star is relatively compact, the accreted wind can be observed by its accretion luminosity in a symbiotic system.  Accretion rejuvenates the companion star, leaving it hotter and bluer than it otherwise would be for its age. When the binary system is in a stellar cluster, such stars are seen as blue stragglers (Section~\ref{sec:binary-classes} and \ref{ssec:binary-stars-that-evolve-in-clusters}).

The direct descendants of intermediate-period blue stragglers are thus the barium stars (with metallicity $[\mathrm{Fe}/\mathrm{H}]\sim 0$), CH stars ($[\mathrm{Fe}/\mathrm{H}]\sim -1$) and carbon-enhanced, metal poor (CEMP) stars ($[\mathrm{Fe}/\mathrm{H}]\lesssim -2$). Recent observations have shown that the CEMP stars are truly equivalent to CH and barium stars \citep{Starkenburg2014}, and that the amount of accreted material is a function of orbital period \citep{Merle2016}.
 The properties of these systems allow direct tests of uncertain physical processes such as wind accretion efficiency, wind Roche lobe overflow \citep{Abate2013}, mixing in the accreting star (e.g. thermohaline mixing; \citealt{Stancliffe2013}) and nucleosynthesis in stars that died many billions of years ago. 
 
Short-period barium stars are often in eccentric systems despite the fact that the binaries have close enough orbits that they should circularise quickly and enter Roche-lobe overflow. These binaries avoid such mass transfer and circularisation, suggesting that our basic theory (as described above) is incorrect. Mechanisms to increase binary eccentricity, such as circumbinary disk interactions and episodic mass transfer, have been tested \citep{Vos2015}, but no clear picture of which process is responsible has yet emerged.

\section{CURIOUS AND COMPLEX PHENOMENA WITH A POSSIBLE BINARY (OR PLANETARY) TWIST}
\label{sec:interesting-and-curious-binary-phenomena}

\subsection{Stripped stellar cores and mergers}
\label{ssec:stripped-stellar-cores-and-mergers}
\begin{figure*}
\centering
\includegraphics[width=0.7\textwidth]{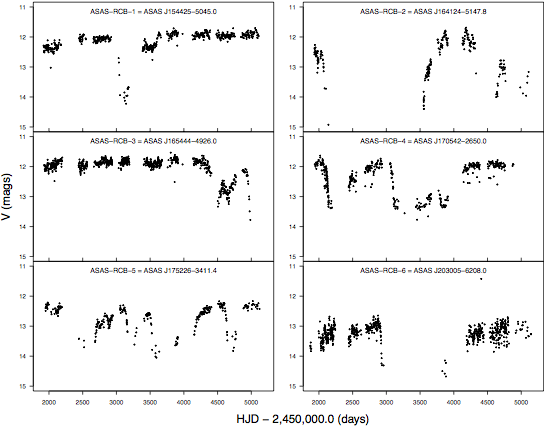}
\caption{{\it All Sky Automated Survey} (ASAS-3) light curves of likely-merger products R~Coronae Borealis Stars ASAS-RCB-1 to ASAS-RCB-6 showing their characteristic, dramatic and random dust obscuration events (Section~\ref{ssec:stripped-stellar-cores-and-mergers}). It is not known what the relationship between the merger history and the dust production properties are. {\it Credit: image adapted from figure 6 of \citet{Tisserand2013}}.}
\label{fig:rcb-tisserand13-fig6}
\end{figure*}

RGB stars have compact, inert helium cores surrounded by a hydrogen-burning shell. Without the hydrogen envelope to fuel the shell, burning stops. In stars with cores of mass less than about $0.45$~\msun, the core then simply cools and forms a helium WD. Some 10 per cent of WD\rgi{s} are made of helium \citep{Liebert2005} and perhaps {\it all} helium WDs are made in binary stars.

If the stellar core has a mass exceeding \rgi{about 0.45~\msun}\ when the envelope is stripped from the RGB star, the core is sufficiently hot and dense that it will ignite in a subsequent phase of helium burning \citep{Han2002}. These stars are known as subdwarf-B and subdwarf-O stars (Table~1) because they have surface temperatures in excess of about 20,000~K \citep{Heber2009}. Most are in binary systems \citep{Jeffery1998,Maxted2001} some of which are eccentric, even though theory suggests they should be circular \citep{Vos2012}. This is a problem comparable to the mystery of the eccentric barium stars (Section~\ref{ssec:wind-polluted-stars}). The single subdwarf B stars may be merged helium WDs \citep{Zhang2012} or simply have companions that we cannot see. 

The merging of two stars is apparently quite a common end point to mass transfer. In addition to the sub-dwarf O and B stars described above, the merged main-sequence stars exhibit properties quite unlike their single star counterparts \citep{Glebbeek2009,Glebbeek2013}. Many blue stragglers may be main-sequence mergers, especially in stellar clusters (\citealt{Hurley2001}; Sec.~\ref{ssec:binary-stars-that-evolve-in-clusters}). 

If a type of star is {\it always} single, it probably forms only when a binary merges. A classic example is the core-helium burning, R-type carbon stars (Table~1). This stellar class comprises stars that are not evolved enough to present carbon at their surface, but equally cannot have accreted carbon from a non-existent companion star \citep{McClure1997}. A helium WD merging with a red giant, itself with a helium core, followed by mixing during helium ignition in a rapidly rotating star, may be the answer.  Population synthesis models predicted that there are sufficient common-envelope mergers to explain these stars \citep{Izzard2007}, and subsequent detailed modelling confirmed that in some cases this may indeed be the case (\citet{Piersanti2010,Zhang2013}. The former study adopted a two-pronged approach combining 3D hydrodynamics and 1D implicit codes; see Section~\ref{sssec:modelling-binary-interactions}). The FK Comae stars are likely post common-envelope mergers, and p!
 erhaps the progenitors of the R stars, because they are red giants that are spinning \rgix{very} rapidly  \citep{Welty1994,Ayres2006}. 

R Coronae Borealis stars (Table~1) \rgi{are} hydrogen-deficient, post-AGB\rgi{,} supergiant pulsators that suffer \rgix{random,} deep \rgi{lightcurve} declines \citep[][Fig.~\ref{fig:rcb-tisserand13-fig6}]{Clayton1996}. \rgi{They are} thought to be merger products\rgix{,} primarily because of the presence of elevated quantities of $^{18}$O \citep{Clayton2007}, \rgi{which can be made} under merger conditions. Models of the mergers have been carried out with 3D hydrodynamic codes \citep{Staff2012}, which in turn were used as inputs to 1D stellar structure codes \citep{Menon2013} to determine the nucleosynthetic properties of the merger. The relationship between the dust activity and the merger past remains unclear \citep{Bright2011}.

The prevalence of close binaries among massive stars \citep[][Section~\ref{ssec:the-binary-fraction}]{Kiminki2012,Sana2012} leads us to the inevitable conclusion that many will interact and merge \citep{deMink2014}. Should a sample of O stars be selected for single stars, as is typical in constructing observing surveys, up to half of these stars are likely to have once been binaries. If they have undergone mass transfer, their stellar structure, internal rotation profile and nucleosynthetic history are likely to be quite different to a single star of equivalent mass. The number of these objects can be constrained by direct comparison with stellar clusters \citep{Schneider2014}, and modern binary population synthesis models agree remarkably well with observed main-sequence stellar mass functions {\it if} binaries are included. The most massive stars, such as the $\sim$320~\msun\ R136a1 \citep{Crowther2010}, are also quite likely to be binary-star mergers \citep{Schneider2014}!
 . This said, a recent investigation of the R136a cluster by \citet{Crowther2016} favours a scenario where not all very massive star are merger products, leaving open the question of how to form such stellar monsters.

\subsection{Polluted white dwarfs}
\label{ssec:polluted-white-dwarfs}
\begin{figure*}
\centering
\includegraphics[width=0.7\textwidth]{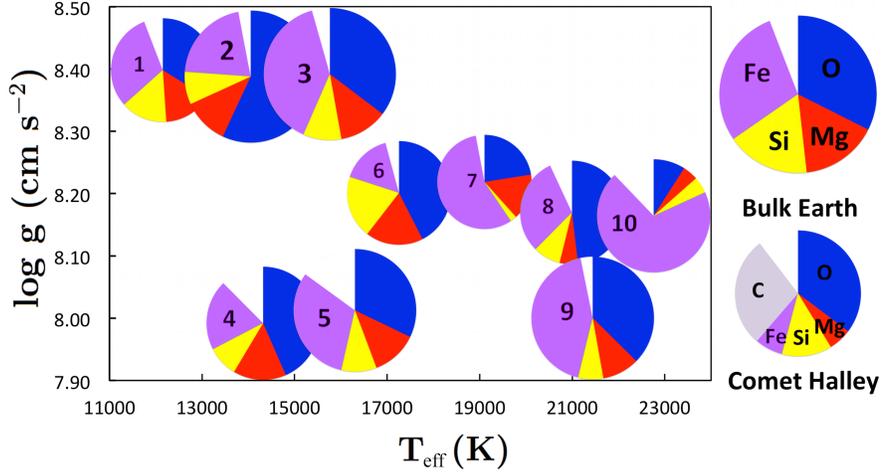}
\caption{Compilation by \citet{Xu2014} of all polluted white dwarfs (Section~\ref{ssec:polluted-white-dwarfs}) with measured abundances for O, Mg, Si, and Fe. The abundances are always the dominant elements in a variety of extrasolar planetesimals, resembling bulk Earth. The abscissa marks white dwarf effective temperature, the ordinate their surface gravity. The size of each pie correlates with the accretion rate. Hydrogen-dominated white dwarfs: 1: G29-38 \citep{Xu2014}, 7: PG 1015+161, 8: WD 1226+110, 9: WD 1929+012, 10: WD 0843+516 \citep{Gaensicke2012}; helium-dominated white dwarfs: 2: WD J0738+1835 \citep{Dufour2012}, 3: HS 2253+8023 \citep{Klein2011}, 4: G241-6, 5: GD 40 \citep{Jura2012}, 6: GD 61 \citep{Farihi2011,Farihi2013}. All white dwarfs except 3 and 4 have a dust disk. Bulk Earth: \citet{Allegre2001}. Comet Halley: \citet{Jessberger1988}. {\it Credit: figure adapted from figure 18 of \citet{Xu2014}.}}
\label{fig:Xu+14-wd}
\end{figure*}

The class of WDs known as DZ includes stars with prominent metal lines in their spectra \citep{vanMaanen1971,Weidemann1960}. The timescale\rgi{s} for settling of metals in the atmospheres of WDs is short compared to the cooling time of the WDs. In many of these objects it is therefore difficult to explain the presence of metals above the photosphere. Between a quarter and half of all WDs exhibit some degree of pollution \citep{Zuckerman2003,Zuckerman2010,Barstow2014,Koester2014}. The hottest WDs ($T \gtrsim$20\,000~K) can achieve metal levitation \citep{Chayer1995,Chayer2014,Koester2014}, but the cooler \rgi{WDs} must have recently accreted the metals \citep{Koester2009}. Early explanations of the pollution phenomenon included accretion from the interstellar medium \citep{Aannestad1985,Sion1988}, but it was not clear why some of the DZ WDs have helium atmospheres, \rgi{because} gas accreted from the interstellar medium would be mostly hydrogen.

\rgi{\citet{Zuckerman1987}} interpreted G29-38, a WD with a prominent infrared excess, as having a brown dwarf companion. Later, a certain number of DZ WDs were discovered to have similar infrared excess flux and a better interpretation was that these WDs were instead surrounded by small dusty disks, \rgi{inside their} Roche limits.

\citet{Alcock1986} was \rgi{the} first to make the connection between the polluted WDs\rgix{,} and the accretion of an asteroid or comet onto the WD. The arrival of an asteroid or comet at the Roche limit of the WD would \rgi{result} in it being pulverised, making a disk that would be visible at infrared wavelengths. Material from the disk \rgi{is} then accreted onto the surface of the WD, polluting the star. To date approximately 30 such disks have been discovered, constituting approximately 1-3 per cent of the studied WDs \citep{Farihi2009}. Occasionally, such WD debris disks are accompanied by a gaseous component, as is the case, for example, in SDSS~J0845+2257 \citep{Wilson2015}. 

Abundance analyses of DZ WD surfaces \citep[e.g.,][]{Dufour2012,Xu2014,Wilson2015} concluded that the observed elemental abundance distribution is similar to that of an asteroid with the average abundance of bulk Earth material (Fig.~\ref{fig:Xu+14-wd}). Subsequent analyses \rgi{claimed} to have \rgix{even} detected patterns produced by asteroids of different compositions, such as those that have differentiated, or by water-carrying asteroids \citep[in GD~61;][]{Farihi2013}.

Questions do arise as to why the metal abundance patterns in some DZ WDs cannot always be explained by accretion, as is the case, for example, for SDSS~J0845+2257, where the carbon abundance is too high to derive from accreted bodies and may be indigenous to the WD \citep{Wilson2015}. 

\rgi{The central star of planetary nebula NGC~6543, the Helix, has a 24-$\mu$m excess caused by the presence of a disk which could derive from disrupted Kuiper belt objects \citep{Su2004}.  This disk may also be a left over from  processes that took place during the AGB, as is likely the case for other central stars of planetary nebulae \citep{Clayton2014} and post-AGB stars \citep{VanWinckel2009}.} 

 The DZ WDs can therefore best be explained if they interacted with an asteroid, implying that 20-30 per cent of all WDs have preserved parts of their planetary systems. This not only indicates that asteroid families commonly survive stellar evolution, but that an undetected,  perturbing planet must exist at large distances from WDs in many cases.

\subsection{The origin of non-spherical planetary nebulae and the unexplained bright edge of the planetary nebula luminosity function}
\label{ssec:the-origin-of-non-spherical-planetary-nebulae}

The debate over what generates non-spherical planetary nebulae (PNe) continues \citep{DeMarco2009b,Kwitter2014}. On the theoretical front, there is still no viable quantitative theory to explain how single stars form highly non spherical PNe, although there could be ways to form mildly elliptical shapes \citep[e.g.,][]{Soker1999}. Single stars cannot sustain the interplay of rotation and magnetic fields that can alter the geometry of the AGB super-wind from a spherical, or almost spherical distribution \citep{Soker2006,Nordhaus2007,GarciaSegura2014}. Naturally only a small fraction of PN can derive from interacting binaries because only a small fraction of binary systems has the appropriate orbital period to interact on the AGB (Section~\ref{ssec:the-period-distribution}). Yet 80 per cent of all PN are non spherical \citep{Parker2006}, only some of which could be explained at the moment by single stars.

\begin{figure}
\centering
\includegraphics[width=0.45\textwidth]{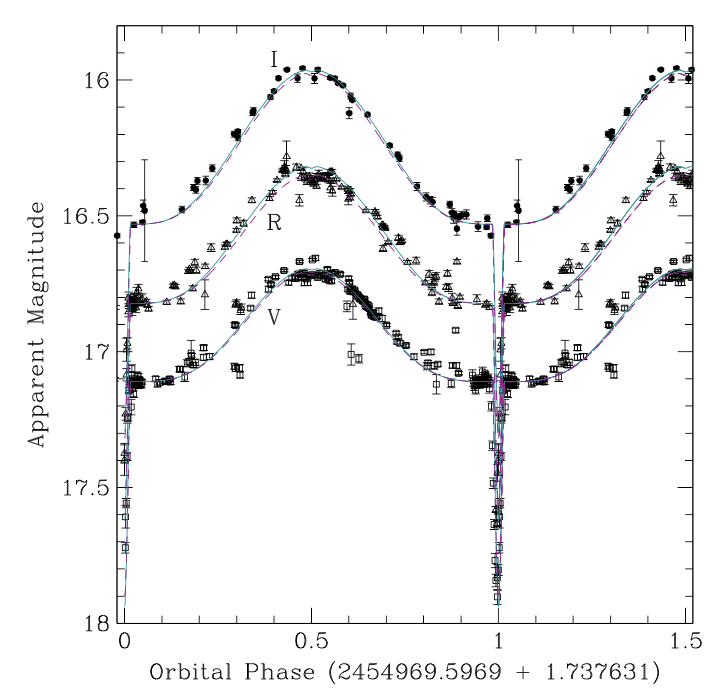}
\caption{Observed $V$, $R$, and $I$ phase-folded light curves of the post-common envelope central binary star of planetary nebula  HaTr 4 for a period of 1.74~days. The variability is due to a combination of irradiation of the main sequence companion by the hot central star as well as eclipses. Such close binaries comprise at least 15~per cent of all central stars of planetary nebula. The solid and dashed lines correspond to two models using a Wilson-Devinney code using different parameter sets, as described in \citet{Hillwig2016}. {\it Credit: image adapted from figure 1 of \citet{Hillwig2016}}}
\label{fig:Hillwig+16-Fig1}
\end{figure}

Alongside this problem, there is a host of additional key observations, which any comprehensive theory of PN formation must be able to explain:

(1) At least 15-20 per cent of \rgi{PNe} have post-common envelope central stars (Section~\ref{sssec:modelling-binary-interactions-the-important-case-of-the-common-envelope-interaction}) in their centres, detected by light variability due to irradiation, ellipsoidal effects or eclipses in the close binaries \citep[see Fig.~\ref{fig:Hillwig+16-Fig1} and Section~\ref{sec:binary-classes}; ][]{Bond2000,Miszalski2009}. These derive from those close main sequence binaries with progenitor primary masses between 1 and 2~\msun, companion masses $\lesssim$1~\msun, with an orbital separation \rgi{shorter} than 2-3 times the maximum AGB stellar radius of the primary of about about 600~\rsun \citep{Villaver2009,Mustill2012,Madappatt2016}. The expected fraction of post-CE central strs of PN is of the order of 5-10 per cent (\citealt{Han1995,Nie2012,Madappatt2016}) lower than the lower limit imposed by observations. This is not explained by current theory.

\begin{figure*}
\centering
\includegraphics[width=0.7\textwidth]{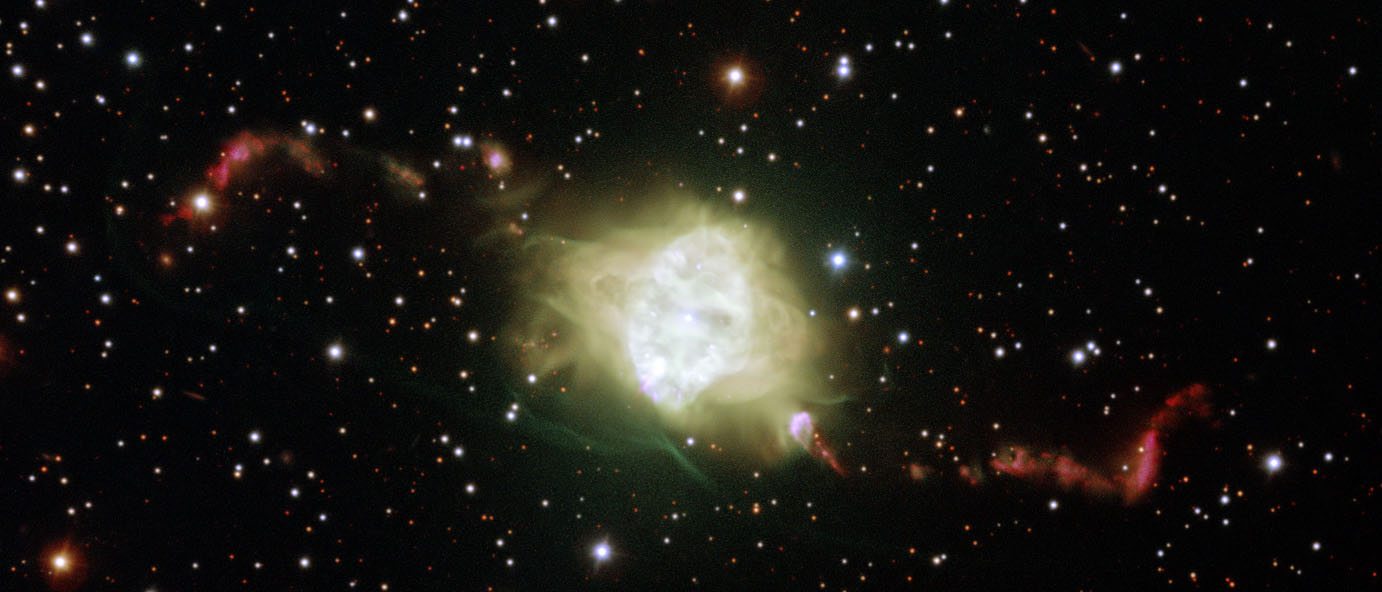}
\caption{The planetary nebula Fleming~1, with its prominent jets, was ejected during a common envelope (Section~\ref{sssec:modelling-binary-interactions-the-important-case-of-the-common-envelope-interaction}) interaction between an AGB star and its companion. The core of the AGB star and the companion are today at the core of the nebula (Section~\ref{ssec:the-origin-of-non-spherical-planetary-nebulae}). {\it Credit: image from figure 2 of \citet{Boffin2012}.}}
\label{fig:fleming1}
\end{figure*}

(2) Post-common envelope binary central stars are preferentially found inside bipolar nebulae, although some post-common envelope central stars are in elliptical PN \citep{DeMarco2009b,Miszalski2009b,Hillwig2016b}, often with jets (Fig.~\ref{fig:fleming1}). 
The scale heights of spherical and bipolar PN \rgi{are} quite different (130 vs. 325~pc; \citealt{Corradi1995}), which points to a larger progenitor mass \rgi{in} bipolars. There is also an association of bipolar PN with those of type~I (N/O$>$0.8; \citealt{Kingsburgh1994}), which must derive from progenitors with a mass larger than 3-5~\msun\ \citep[][]{Karakas2009}. Only stars with initial mass larger than 5~\msun\ can make type I PN. However, there could be mixing processes that allow stars with initial mass as low as 3~\msun\ to develop \rgix{the} type I abundances. It is hard to reconcile the relatively large percentage of type I PN ($\sim$20 per cent; \citealt{Kingsburgh1994}), with the initial mass function, that indicates that the fraction of stars more massive than 3~\msun\ is of the order of few per cent. It is even harder to understand the association of post-CE PN with a more massive population, although \citet{Soker1998} noted that the fact that more massive mai!
 n sequence stars have binary companions more often (Section~\ref{sec:main-sequence-binary-stars}), and that they grow to larger radii, would promote a  correlation between type I PN, post-CE PN and bipolarity.

(3) \citet{Nie2012} used a binary population synthesis model calibrated to the fraction of giant stars that exhibit the sequence E phenomenon (Section~\ref{ssec:types-of-binary-mass-transfer}), \rgix{(due to the presence of a close companion),} to predict how many \rgi{PNe} derive from a binary interaction. They concluded that \rgi{49-74 per cent of PNe come} from non-interacting binaries and single stars. However, they also predicted a fraction of single central stars in the range 3-19 per cent, which cannot be reconciled with the much larger multiplicity fraction of the progenitor population (50$\pm$4 per cent for Solar-type stars; $M_{\rm{MS}} = 1-1.3~$\msun; \citealt{Raghavan2010}, Section~\ref{ssec:the-binary-fraction}). A way to reconcile these numbers with data from main sequence stars is to assume that not all 1-8~\msun\ stars make a visible PN. If AGB stars that interact with a companion made a brighter PN, then the detected fraction of post-interaction PN would be !
 inflated.

(4) The bright edge of the PN luminosity function is almost invariant and\rgi{,} if one allows for small metallicity-dependent corrections, it predicts the distance to external galaxies with excellent precision for both old ellipticals and \rgi{young} spiral galaxies \citep{Ciardullo2010}. This indicates a ubiquitous population of relatively massive central stars in all galaxies\rgi{. This disagrees with} the prediction that \rgi{PN in} old elliptical galaxies \rgix{should} have lower-mass central stars. \citet{Ciardullo2005} \rgi{argued} that the data is consistent with PN from blue straggler stars (Section~\ref{ssec:binary-stars-that-evolve-in-clusters}) populating the bright end of the luminosity function.

\subsection{The effects of binary interaction on the population of Wolf-Rayet stars and luminous blue variables }
\label{ssec:luminous-blue-variables-and-wolf-rayet-stars}

The fact that \rgi{about} 70 per cent of massive stars interact with a companion  (Section~\ref{sec:main-sequence-binary-stars}) reinforced the suspicion that previously known massive star phenomena may have a binary origin, at least in a fraction of the objects. Here we concentrate in particular on the Wolf-Rayet phenomenon (Section~\ref{sssec:wolf-rayet-stars}), luminous blue variables \rgi{(LBVs; Section~\ref{sssec:luminous-blue-variables})} and core collapse supernovae (see also Section~\ref{ssec:type-I-supernovae}).

\subsubsection{Wolf-Rayet stars}
\label{sssec:wolf-rayet-stars}

Wolf-Rayet (WR) stars \citep{Wolf1867} are rare, luminous stars with strong emission lines, occasionally with P-Cygni profiles of helium, nitrogen, carbon and oxygen indicating strong, mass-losing winds. WR stars dominated by nitrogen lines are called "WN". The "WC" and "WO" types have emission lines of, predominantly, carbon and oxygen, respectively. It is thought that they form a sequence in that the earlier phase, the WN, gives rise to the later phase, WC/O once the strong winds have eliminated the nitrogen-rich layer \citep[for a review see][]{Crowther2007}. 

The winds of WR stars are line-driven as was discovered once UV-spectroscopy of massive stars became available \citep{Morton1967}. The original explanation of this rare phenomenon leaned towards a binary interpretation, where the companion strips the mass off the massive star by Roche lobe overflow \citep[e.g.,][]{Paczynski1966}. However, the realisation that WR stars have high mass loss rates, up to $10^{-4}$~\msun~yr, as well as a series of population studies that reconciled the relative numbers of O and WR stars \citep{Massey2003}, provided a reasonable, single-star explanation for the WR phenomenon. Later, the understanding that WR winds are clumped lead to a downward revision of the the mass-loss rates deduced from UV and optical observations \citep{Vink2005} and, once again, the interpretation of the WR phenomenon included, at least in part, the effects of binary interactions.

There are several cases in which a WR star is known to have a binary companion. The binary fraction in the massive WR population is 40 per cent both in the Galaxy and in other populations such as the Small Magellanic Cloud \citep{Foellmi2003}. This is somewhat contrary to the expectation that a lower metallicity would reduce stellar wind mass loss rates and leave only binary interactions to strip stars of their hydrogen, something that would have driven the Small Magellanic Cloud WR binary fraction up. It is also expected that in a binary, even lower mass stars may develop the WR phenomenon. However, \citet{Shenar2016} found that in the Small Magellanic Cloud WR binaries have masses in excess of the limit above which single stars should be capable of entering a WR phase at those metallicities. Clearly mass-loss has a large impact on the evolution of the star and its observed quantities, but just what the interplay of binarity and mass-loss is on the WR phenomenon remains at !
 this time hard to pinpoint.

\citet{Langer2012} argues that 1) the detected population of WR stars is not particularly impacted by binary interactions, in other words we can be reasonably sure that the current WR stars are not the product of mergers and 2) that there is a large population of low mass, low luminosity WR stars that is thus far undetected. Their progenitors are massive Algols where the donor can be an O or at most an Of/WN star \citep{Rauw1999}. The donor will likely develop into a WR star when it loses more mass to the companion in what \citet{Langer2012} calls Case AB mass transfer. Their progeny would be the Be/X-ray binaries \citep{LiuQZ2006}. 

\begin{figure}
\centering
\includegraphics[width=0.46\textwidth]{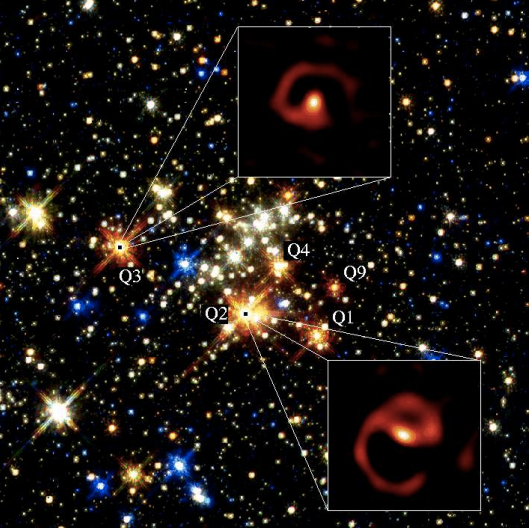}
\caption{Multiwavelength {\it Hubble Space Telescope}, {\it Near Infrared Camera and Multi Object Spectrograph} near-infrared image of the Quintuplet cluster (for details of the image see \citealt{Figer1999}). The five red stars are labelled according to the nomenclature of \citet{Moneti1994}. All of them are "dustars", dust-producing, binary  Wolf-Rayet stars (Section~\ref{sssec:wolf-rayet-stars}). Inset images of Q2 and Q3 recovered with {\it Keck} telescope speckle interferometry are overlaid, with graphical indication showing the relative scaling between the {\it Hubble} and {\it Keck} images. {\it Credit: image adapted from figure S1 of \citet{Tuthill2006}.}}
\label{fig:quintuplet}
\end{figure}

Aside from WR stars in short period binaries (e.g., SMC~AB6, with a 6.5 day period; \citealt{Shenar2016}), some WR binaries have periods from tens to hundreds of days and some are found to have colliding winds, as observed in X-ray light curves (e.g., WR21a; \citealt{Gosset2016}). Occasionally, dust forms in the wake of these collisions and the WR can be observed as a pinwheel nebula such as WR104, discovered by \citet{Tuthill1999}. Since the prototype, several other similar pinwheel nebulae have been discovered, such as WR98a (\citealt{Monnier1999}), as well as two in the Quintuplet star cluster \citep{Tuthill2006}. This cluster is one of the most massive in our Galaxy and is named after the five mysterious red sources with very high luminosity, which have been interpreted both as young and evolved stars \citep{Okuda1990}.  All five  sources were resolved by the {\it Keck} telescope using speckle interferometry \citep{Tuthill2006} and two of these sources, with the largest !
 sizes, are superimposed on a {\it Hubble Space Telescope} image of the cluster  in Fig.~\ref{fig:quintuplet}. 

It is likely that a number of other WR stars with dust implied by high IR fluxes or with colliding winds implied by non-thermal radio emission (for example WR104, WR98a, as well as WR48a WR112 and WR140; for a summary see \citealt{Monnier2007}) may be such pinwheels but either too far or at a non-favourable inclination to be resolved (e.g., WR112; \citealt{Monnier2007}). While these binaries may be too wide for violent phenomena to occur, an important aspect of their geometry is that their wind-wind collision zones seem to promote the manufacture of carbon dust, despite the hot environment, sometimes at rates as high as $10^{-6}$~\msun~yr$^{-1}$ \citep[][]{Williams1995}. It is not known how many dust-making WR stars ("dustars") exist in the Galaxy today, but numbers such as 100-1000 are not unlikely based on new surveys \citep{Shara2009,Shara2012b} that are finding large numbers of the cooler, WC9 type WR stars near the Galactic centre. If this were the case, this type of WR!
  binary would produce dust at rates commensurate to the classic dust producers, such as red supergiants, AGB stars, planetary nebulae and supernovae \citep{Marchenko2007,Drain2009}.

\subsubsection{Luminous blue variables}
\label{sssec:luminous-blue-variables}

LBVs have for a long time been interpreted as massive stars at the transition between the end of hydrogen burning and the start of core helium burning that undergo eruptive phases of mass loss. In a scenario laid out first by \citet[][]{Conti1978}, O stars go through strong mass-loss rates turning them into Wolf-Rayet stars of class WNH \citep{Smith2008}, before an LBV eruption frees them of all hydrogen \rgi{and turns} them into WC/WO Wolf-Rayet stars and eventually type Ib and Ic supernovae \citep[e.g.,][]{Crowther2000}. At first it was thought that the main mechanism by which LBVs shed large amounts of mass would be line-driven winds \citep{Humphreys1994,Lamers2002}. In this scenario, the LBV outburst is due to an increase in the bolometric luminosity together with a decrease of the mass via mass loss, which leads to a high ratio of luminosity to mass, which in turn brings the star close to the Eddington limit.   
However \citet{Smith2006} showed that a more likely mechanism is a super-Eddington wind driven by continuum radiation pressure.  

Several problems affect the standard LBV scenario and the evolution of massive stars in general. The LBV S~Doradus had an outburst with a measured  mass-loss rate far below that needed to explain the expanding pseudo-photosphere envisaged by classical LBV theory \citep{deKoter1996,Groh2009a}. Nor did the luminosity changes observed happen at constant bolometric luminosity \citep{Groh2009a}. 
The brightness outburst in S~Doradus seemed rather driven by a "pulsation" of the envelope \citep{Graefener2012}. Similarly, light echo spectra of the Great Eruption of $\eta$~Carinae \citep{Rest2012} are inconsistent with a pseudo-photosphere and more in line with the spectra of transients NGC~4990-OT and V~838~Mon \citep[][see also Sec.~\ref{ssec:intermediate-luminosity-optical-transients}]{Smith2016b}, that are suspected binary mergers. 

Further problems \rgi{arise because} the most likely progenitors of some type IIn supernovae (hydrogen-rich supernovae with narrow lines) have LBV-like mass-loss rates and in four cases progenitors are known, e.g., SN 1961V \citep{Smith2011c,Kochanek2011}, 2005gl \citep{GalYam2009}, 2010jl \citep{Smith2011c} and 2009ip (\citealt{Smith2010b}, \citealt{Foley2011}; see also  \citet{Smith2014} for a review). However, in the Conti scenario LBVs do not explode as type~II supernovae, rather they \rgi{spend} 0.5-1~Myr as Wolf-Rayet stars, which then explode as \rgi{type Ib or Ic supernovae}.  

Approximately a dozen (eruptive) LBVs are known in the Galaxy and Magellanic Clouds \citep[e.g.,][]{Clark2005} with another dozen known in external galaxies. \citet{Smith2015} found that LBV stars are statistically more isolated than O and Wolf-Rayet stars in the Galaxy and in Large Magellanic Cloud  clusters, indicating that LBVs cannot in all cases be a phase in the evolution of O and Wolf-Rayet stars. They proposed alternative binary scenarios for at least a fraction of the objects.

\begin{figure}
\centering
\includegraphics[width=0.45\textwidth]{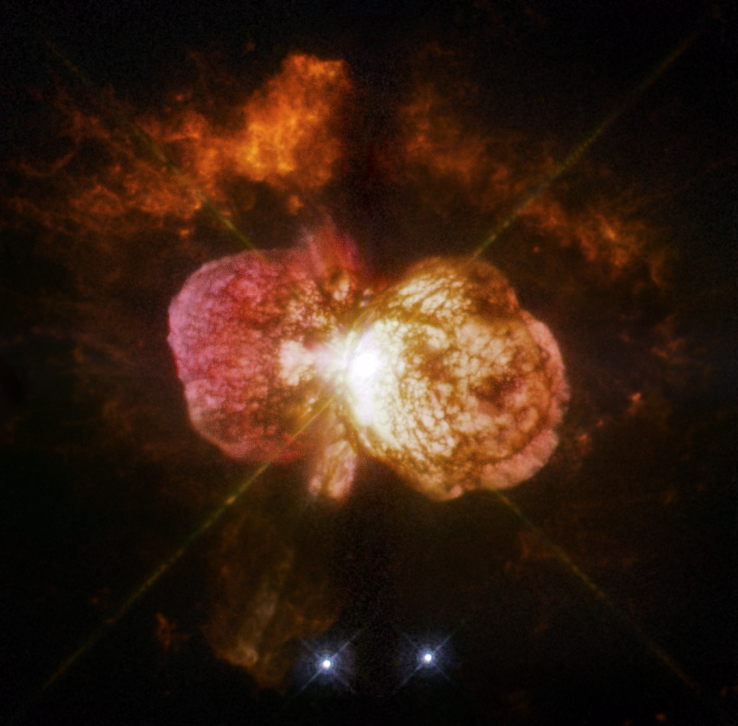}
\caption{{\it Hubble Space Telescope}, {\it Wide Field and Planetary Camera 2} images of gas ejected during the Great Eruption of the luminous blue variable $\eta$~Carinae (Section~\ref{sssec:luminous-blue-variables}). False colour in 5 optical bands. The image is  approximately 40 arcseconds on a side. North is towards the bottom left corner, east is towards the bottom right corner. {\it Credit: image courtesy of the Hubble Site, associated with press release STScI-2009-25}.}
\label{fig:etacar}
\end{figure}

It was \citet{Gallagher1989} who first discussed the LBV phenomenon in connection \rgi{with} binary \rgi{stars}. 
Many papers \rgix{have later} discuss\rgix{ed} the famous $\eta$~Carinae eruptions (Fig.~\ref{fig:etacar}) arguing for and against a binary interpretation. A B-type companion was announced in 2005 \citep[][see also \rgi{\citealt{Soker2001c}} and references therein]{Iping2005}, but it later transpired that the observations that were interpreted as a companion detection could have been explained by alternative effects \citep[e.g.,][]{Martin2006}. The 5.5 year spectroscopic cycle \citep[e.g.,][]{Zanella1984} of emission lines periodically increasing in strength, also seen as an X-ray brightening \citep{Ishibashi1999}, tends to be explained as a companion in an eccentric orbit  where the X-ray luminosity increases near periastron due to colliding winds and the spectroscopic event is due to mass ejection at periastron \citep[e.g.,][]{Mehner2010}. However, a direct detection of the companion is yet to be accomplished. 

\citet{Iben1999} suggested that the Great Eruption of $\eta$ Carine was due to a merger in a triple system, but it is difficult to understand further eruptions in the same system. \citet{Soker2004} envisaged how the Great Eruption could be interpreted as a binary interaction at the time of periastron and even extended this interpretation to some "supernova impostors" (see Section~\ref{sssec:supernova-impostors}). Accretion at periastron passage would power jets that eventually form the Homunculus, the bi-polar structure that is so well known today (Fig.~\ref{fig:etacar}). 

However, some aspects of this presumed interaction are inconsistent with observations \citep{Smith2011d}. In particular one question is the distance of the periastron passage compared to the photospheric radius and the Roche lobe of the primary. \citet{Staff2016} showed that any mass transfer between the primary and the secondary at periastron passage shortens the period of the binary leading eventually to a common envelope phase and possibly a merger. This would mean that the 5.5 year orbit would not be stable as is instead observed.

\rgi{It} is likely that there are multiple classes of LBVs\rgix{,} and that the proximity of a companion \rgi{plays} a major role in \rgi{many} LBVs, particularly when \rgi{the stellar} mass is too low to exhibit \rgi{single-star} instabilities. 
\rgix{Even in the more massive cases, it is possible that the eruption is due to an interplay of the massive star behaviour with the companion action, while it is still possible that some of the most massive stars may have instabilities of their own accord, though perhaps not all the times.}
Ultimately, just the range of parameters that \rgi{drive} binary evolution compared to those that regulate a single star, would help expain a broad range of behaviours, \rgi{such that} identifying specific classes \rgi{is} difficult. In \rgi{Sections~\ref{ssec:type-I-supernovae}} and \ref{sec:transients} \rgi{we discuss the related phenomena of supernovae} and supernova impostors, emphasising how upcoming transient surveys will give us a sufficient number of objects to start identifying broad behavioural classes.

\subsection{Hydrogen-deficient, type I supernovae}
 \label{ssec:type-I-supernovae}
  
Supernovae of type I are explosions, presumably of stars, that show no evidence for hydrogen \rgi{in their spectra}. These can be subdivided into the silicon-rich Ia, \rgi{and} types Ib and Ic, which have no, or weak silicon lines in their spectra, but have helium (Ib) or \rgi{do} not (Ic). The modern consensus is that the type Ia supernovae are thermal runaway explosions of carbon-oxygen WD \citep[e.g.,][]{Hillebrandt2000}, while Ib and Ic supernovae are helium stars that undergo core collapse \citep[e.g.,][]{Smartt2009}. The mechanisms for explosion are quite different, but binary stars are related to both.

\subsubsection{Type Ia supernovae}
 \label{sssec:type-Ia-supernovae}
 
Type Ia supernovae are famous for having provided the standard candles required by cosmologists to deduce that the Universe is expanding \citep{Leibundgut2001}.
They are not true standard candles, rather they are {\it standardizable}, meaning that their maximum brightness correlates with the width of their lightcurves \citep{Phillips1993}, at least in local \rgi{type Ia supernovae}.
Thus, given their lightcurve width, their intrinsic brightness can be calculated, and hence their distance. 

Despite their successful use as cosmological tools, we do not know \rgi{for sure} what causes a \rgi{type Ia supernova}. The most likely scenario is that \rgi{type Ia supernovae} are exploding carbon-oxygen WD with masses near the Chandrasekhar mass limit of about~$1.4\,$\msun\ \citep{Wang2012}. WDs are no longer undergoing nuclear fusion and their gravitational collapse is resisted by electron degeneracy pressure.
If such a WD accretes mass in a binary-star system, the central density and temperature increase and, eventually, core carbon ignites. The degenerate nature of the star leads to a thermonuclear runaway and to the disruption of the entire star. It is necessary to invoke common envelope interactions (Section~\ref{sssec:modelling-binary-interactions-the-important-case-of-the-common-envelope-interaction}) to make stellar systems containing a carbon-oxygen WD in a suitably short orbit for any mass transfer to occur. 

There are many outstanding theoretical problems related to \rgi{type Ia supernovae}. The details of the explosion matter greatly to the nucleosynthetic signature of its remnant. Whether the explosion is a subsonic deflagration, a supersonic detonation or some combination of the two is unclear at present. There are successful 2D and 3D models of type Ia supernovae (e.g. \citealt{Fink2014}), although the explosion triggering is still largely based on simplified physics. 

The total number of systems which explode as \rgi{type Ia supernovae} can be predicted by binary population synthesis models (Section~\ref{sssec:modelling-binary-populations}). While there are many uncertainties involved, not least the problem of common envelope evolution described above, estimates between different research groups \rgi{are remarkably consistent} \rgix{with each other} \citep{Toonen2014}. However, these theoretical estimates are a factor \rgi{of} 4 to 10 lower than observational rates \citep{Claeys2014}. The  problem is that we \rgi{just} do not know which stars explode as type Ia supernovae, or how their progenitor systems form in the first place. 
Double degenerate systems contain two WDs \citep{Nelemans2005} which, if they can merge to form a single WD in excess of the ignition mass, may explode. Such systems are observed, but it remains a great challenge to model their formation. Maybe triple systems offer a solution in some cases, increasing the rate of merging, but it is not clear \rgi{that} there are enough \rgi{systems} in the appropriate parameter space to match the observed \rgi{type Ia supernova} rate \citep{Hamers2013}.

\rgi{Single-degenerate} systems remain candidates for type Ia supernova progenitors, but the evidence is mixed. They involve mass transfer from a giant or sub-giant star to a WD, which increases its mass beyond the limit for ignition. No type Ia supernova contains hydrogen, which would be expected from the majority of donor stars, thus putting this model in doubt or at least rendering it rare. An alternative is helium donors, which are likely significant \citep{Claeys2014}. This said, despite repeated searches for companions, some of which imposed stringent limits on their absence \citep{Schaefer2012},  strong evidence for a main-sequence companion was recently presented by \citet{Marion2016} and a UV signature detected by the {\it Swift telescope} 4 days after the explosion is consistent with supernova ejecta impacting a companion star in the case of iPTF14atg \citep{Cao2015}.

A comparison of different scenarios that may lead to a type Ia supernova explosion can be found in \citet{Tsebrenko2015} who also discuss the fraction of type Ia supernovae that may occur inside of a planetary nebula.

\subsubsection{Type Ib,c supernovae}
 \label{sssec:type-Ibc-supernovae}
 \begin{figure}
\centering
\includegraphics[width=0.4\textwidth]{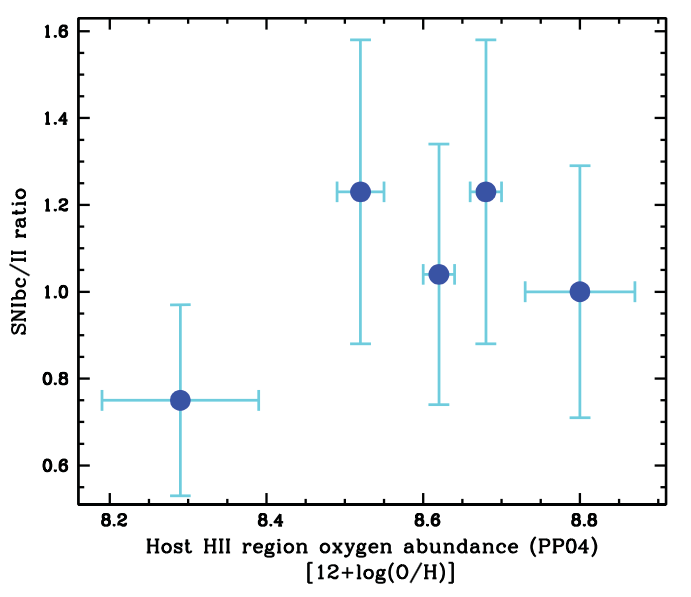}
\caption{The ratio of supernovae type Ib and Ic to supernovae type II as a function of metallicity, measured from the oxygen abundance in HII regions. The lack of a decrease in this ratio with increasing metallicity argues for a mixed origin  for these type of supernovae (Section~\ref{sssec:type-Ia-supernovae}). {\it Credit: image adapted from figure 10 of \citet{Anderson2015}.}}
\label{fig:sn-ratios}
\end{figure}

Type~Ib~and~Ic supernovae are thought to be explosions caused by the collapsing \rgi{cores} of stars that have lost their hydrogen (Ib) and helium (Ic) \rgi{envelopes}.   Because a massive star must lose \rgi{much} of \rgi{its} mass to expose  its helium or carbon--oxygen core, binaries are naturally invoked as a cause \citep{Podsiadlowski1992,Eldridge2008}. However, stellar wind mass loss is significant even in single massive stars \citep{Vink2001} and, perhaps combined with rotation, is sufficient to lead to \rgi{type Ib and Ic supernovae}. The formation of progenitor systems of \rgi{type Ib and Ic supernovae}, the Wolf-Rayet stars, can probably form by both wind mass loss and rotational mixing in rapidly spinning single stars \citep{Yoon2005} as well as binary interactions. 

\rgi{Helium stars with masses in excess of $5$\msun} should have been seen by various supernova progenitor search programmes, but they \rgi{have} not. This may imply that massive helium stars do not explode as supernovae, instead \rgi{they collapse directly} to black holes. \rgi{Lower-mass} helium stars are even more likely to form by of binary interactions because wind mass loss is weaker at lower luminosities and hence lower mass. Searches for progenitors may simply not be sensitive enough to see low mass helium star progenitors of distant supernovae \citep{Yoon2012}. 

\citet{Eldridge2013}  presented an extensive search for the progenitors of type Ib and Ic supernovae in all available pre-discovery imaging since 1998, finding that 12 type Ib and Ic supernovae have no detections of progenitors in either deep ground-based or {\it Hubble Space Telescope} imaging. They showed that the deepest absolute $B$, $V$ and $R$-band magnitude limits are between $-4$ and $-5$. By comparing these limits with the observed Wolf-Rayet population in the Large Magellanic Cloud they estimated statistically that a failure to detect such a progenitors by chance is unlikely. They proposed an alternative that the progenitors of type  Ib and Ic supernovae evolve significantly before core-collapse. 

\citet{Eldridge2013} also reviewed the relative rates and ejecta mass estimates from light-curve modelling of type Ib and Ic supernovae, and found both data sets incompatible with Wolf-Rayet stars with initial masses $>$25~\msun\ being the only progenitors. Finally, they presented binary evolution models that fit the observational constraints and determined that stars in binaries with initial masses $\lesssim$20~\msun\ lose their hydrogen envelopes in binary interactions to become low-mass helium stars. They retain a low-mass hydrogen envelope until $\sim$10$^{-4}$~yr before core-collapse, so it is not surprising that Galactic analogues have been difficult to identify.
The predictions of \citet{Eldridge2013} may have been bourn out in the discovery of a possible progenitor of SN iPTF13bvn \citep{Cao2013} that is consistent with a lower mass helium star, but inconsistent with a Wolf-Rayet progenitor \citep{Eldridge2015}. 

\citet{Crowther2013} concluded that supernovae type Ib and Ic are more frequently associated with HII regions than type II supernovae, pointing to a larger progenitor mass, though he could not differentiate between type Ib and Ic. \citet{Anderson2015} showed that type Ic are more often associated with H$\alpha$ emitting galaxies than type Ib, pointing to a higher progenitor mass for that type. This conclusion is also in line with the study of \citet{Smith2011}, who advocated a mixed origin for the supernovae type Ib and Ic, with single stars able to produce some supernovae type Ic, which would then have a higher progenitor mass. Additionally,  single stellar evolution predicts an increase in the ratio of supernova type Ib and Ic to supernova type II with increasing metallicity, due to higher mass-loss rates at higher metallicity \citep[e.g.,][]{Heger2003,Ibeling2013}. However, an increase of the ratio with oxygen abundance, used as proxy for metallicity, was not observed by !
 \citet[][Fig.~\ref{fig:sn-ratios}]{Anderson2015}, except for the lowest metallicity bin, where a lower ratio was reported. 


In conclusion there is little doubt that binary stellar evolutionary channels account for a substantial fraction and for the diversity of supernova types, though the interplay of single and binary evolutionary channels is likely to increase the complexity of the supernova phenomenon.

\section{TRANSIENTS}
\label{sec:transients}

Transients related to binary stellar evolution are either outbursts, or other periodic or semi-periodic light changes. \rgi{Well-known} transients \rgi{include} cataclysmic variables such as novae and dwarf novae (Table~1), which are observed locally and are very numerous. Other, more rare transients are supernovae, which being more luminous can be observed out to much greater distances. Occasionally, a transient is observed that has an unknown nature. These tend to be studied intensely, e.g., V~838~Mon, with over 300 articles since 2002, but seldom form a new class because of their rarity.
\begin{figure}
\centering
\includegraphics[width=0.4\textwidth]{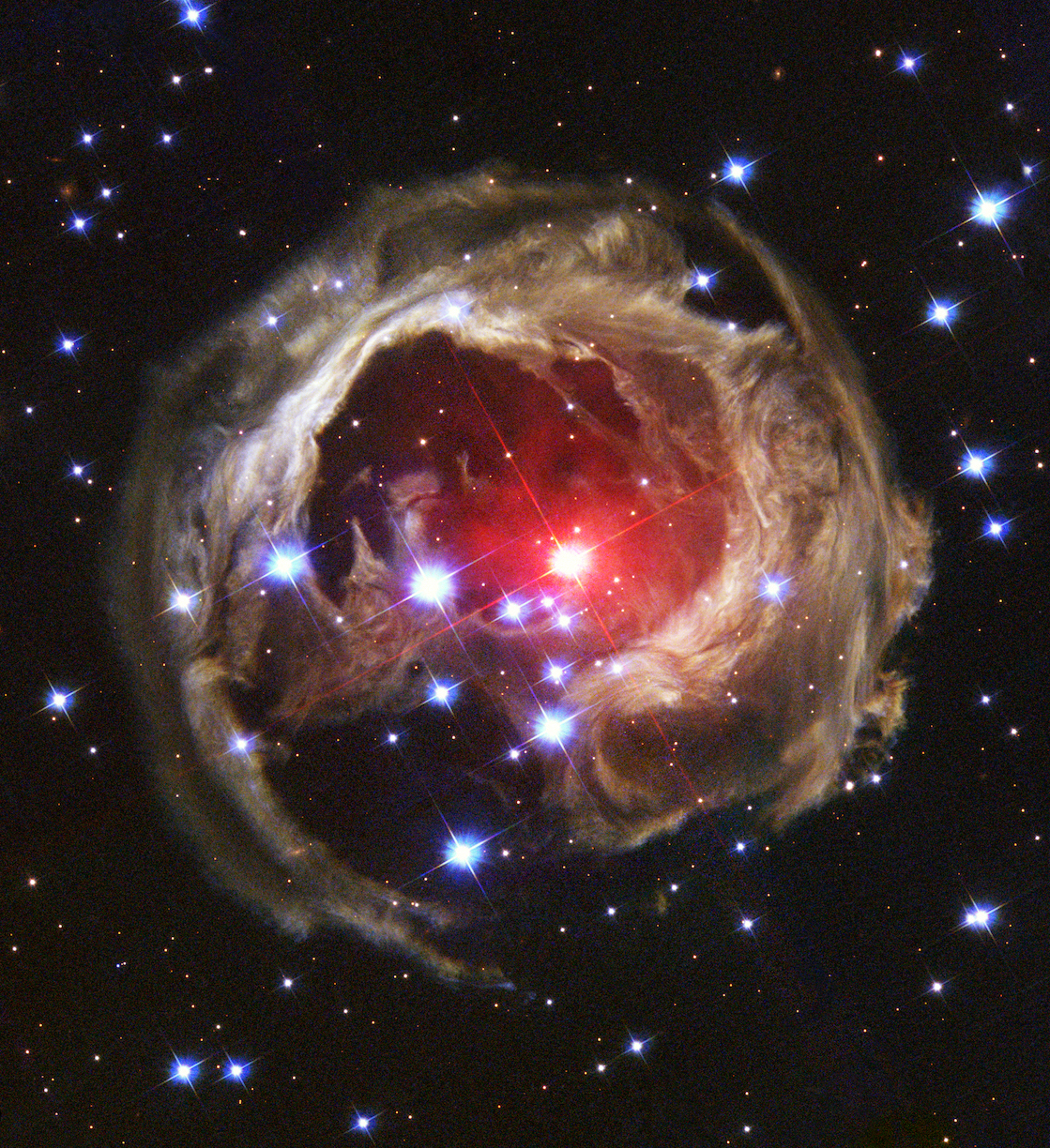}
\caption{{\it Hubble Space Telescope} image of the gap transient V~838~Mon, taken on February 8, 2004 (Section~\ref{sssec:lrn}). This gap transient is thought to be due to a merger of two stars. The dust illuminated by the outburst is not ejected by the object but has an interstellar origin. Composite image constructed using three filters: F435W (B), F606E (V), F814W (I). North is towards the top-left of the image. The image is 2.4 arcmin across or 4.2~parsec at a distance of 6~pc.  {\it Credit: Hubble Space Telescope program 10089, PI Noll}.}
\label{fig:v838mon}
\end{figure}

\subsection{Surveys for transients }
\label{ssec:transient-surveys}

Transient surveys have increased the number of known transients and generated a sufficient number of objects that new classes have been established. Even with modest size telescopes, they have revolutionised the field of transient studies and, because many transients are likely binary phenomena, the field of binaries itself. Even before the latest wave of transient surveys,  which we detail below, supernova searches were detecting transients that do not squarely falling within the supernova classification. For example the {\it Lick Observatory Supernova Search} (LOSS) using a 76-cm robotic telescope \citep{Filippenko2001}. 

Examples of more modern dedicated surveys are the 
{\it Catalina Real-Time Transient Survey} (CRTS; \citealt{Drake2009}), 
the {\it Palomar Transient Factory} (PTF; \citealt{Law2009}), 
{\it Pan-STARRS1} \citep{Kaiser2010}, 
and {\it SkyMapper} \citep{Murphy2009}\rgi{. 
All these surveys are optical in nature and use one and two-meter-class telescopes.}
In addition, surveys designed to detect near Earth objects (e.g., LINEAR, \citealt{Stokes2000}) are also used to detect  transients \citep{Palaversa2016}. There are also transient surveys using non-dedicated telescopes and instruments, usually targeting specific parameter spaces, such as high cadence surveys for fast transients (e.g., {\it Subaru Hyper Suprime-Cam Survey Optimised for Optical Transients}, SHOOT\rgi{;} \citealt{Tanaka2016}, or the {\it High Cadence Transient Survey}, HiTS; \citealt{Forster2014}). 


Upcoming surveys such as the {\it Zwicky transient survey} \citep{Smith2014-ztf}, will survey 3750 square degrees an hour, 15 times faster than its predecessor the PTF. Finally, the {\it Large Synoptic Survey Telescope} (LSST; \citealt{Ivezic2008}), with an 8.4-m mirror and a plan to scan 30\,000 sq. deg. of sky every third night, should detect a huge \rgi{number} of phenomena, many of which are currently unknown. Most will be so faint that they will require follow up by high-demand telescopes such as the {\it James Webb Space Telescope} or a thirty-meter class telescope.  

Other surveys take different strategies. The {\it Public European Southern Observatory Spectroscopic Survey of Transient Objects} (PESSTO; \citealt{Smartt2015}), follows up photometrically and spectroscopically specific transients selected from publicly available sources and wide-field surveys.

At wavelengths other than optical, X-ray transient surveys reveal a range of binary interaction activities, primarily in high mass X-ray binaries with black hole \citep[e.g.,][]{Tetarenko2016} or neutron star \citep[e.g.,][]{Bozzo2008} accretors.
The new {\it SPitzer InfraRed Intensive Transients Survey} (SPIRITS; \citealt{Kasliwal2014}) should detect year-long transients produced by slow in-spiral \rgi{because} of outflow from the second Lagrangian point \citep{Pejcha2016}. It \rgi{can also detect} dust formation in explosive events \citep[e.g., in V~1309~Sco,][]{Nicholls2013} and \rgi{find} transient events \rgi{that have no} optical counterpart. A large number of radio transient surveys have been operational for a long time \citep[e.g.,][]{Williams2013} but the new capabilities of the {\it Australian Square Kilometre Array Pathfinder}, ASKAP (e.g. the {\it Variables and Slow Transient}, VAST, survey) and eventually the {\it Square Kilometre Array} will add a new dimension to the searches \citep[see, e.g.,][]{Metzger2015}.

If we add to these upcoming surveys the new capability of gravitational wave detection (Section~\ref{ssec:gravitational-wave-sources}), we see how their combined power provides us with a new tool to study interacting binaries and connect scattered events into a coherent picture.

\subsection{Gap transients}	
\label{ssec:intermediate-luminosity-optical-transients}
 
The luminosity gap between the faint and numerous novae and the bright but rarer supernovae is being increasingly  filled. Such transients used to be discovered by amateur astronomers (e.g., SN2008S, \citealt{Arbour2008}) or serendipitously, as is the case for M31 RV that erupted in 1988 \citep{Rich1989}. Such discoveries were only sporadically followed up.  The proliferation of new surveys such as the CRTS and PTF has increased the number of gap transients detected. 

We distinguish three types of gap transients following the nomenclature of \citet{Blagorodnova2016}: supernova impostors thought to be non-terminal eruptions in massive stars such as LBVs; intermediate luminosity optical (or red) transients (ILOTs/ILRTs) explained as faint terminal explosions and luminous red novae (LRNs), which are potential stellar mergers. The terminology ILOT is, however, variably used to encompass all gap transients, for example by \citet{Kashi2016}, who also envisaged, but not without controversy \citep{Smith2011c}, a more unified interpretation for the entire class.  The division above is based on interpretation, rather than on observational characteristics. While a classification system should  stay away from interpretation, it is possible that at this time the observational qualities of these transients are still too disparate and the observations too uneven to lead to a proper classification.

\subsubsection{Gap transients: supernova impostors}	
\label{sssec:supernova-impostors}

Searches for supernovae have discovered eruptive events thought to be similar to LBV eruptions, which are too rare to be readily observed in our Galaxy. These have been called supernova "impostors". Supernova impostors are characterised by type IIn spectra with lower peak luminosities than typical core collapse supernovae ($M_V \sim -13$ instead of $\sim -17$; \citealt{VanDyk2000}).

It has been realised in the past decade that there is quite a diversity among the supernova impostors. While some have high luminosity and may derive from high mass stars, some may come from stars with  lower mass progenitors \citep{Prieto2008,Thompson2009} that should not approach the Eddington limit, raising the possibility of alternative pathways to these phenomena, possibly including a binary companion. Some supernova impostors, with sustained high luminosity phases could be powered by an ejection that ploughs into  circumstellar material, transforming kinetic energy into luminosity. However, we do not yet have a model that produces the circumstellar shell in the first place. 

Examples of the supernova impostor class are $\eta$~Carinae, R127 and S Doradus \citep{Walborn2008} in the Large Magellanic Cloud, SN~2009ip \citep{Fraser2013} or UGT 2773 OT2009-1 \citep{Smith2016}. These eruptions happen in dusty environments created by past outbursts. Some impostors have been, in turn, interpreted as massive binary stars. For example the X-ray signature in supernova SN~2010da is consistent with it being a high mass X-ray binary \citep{Binder2011,Binder2016}. Some supernova impostors \rgi{could} be powered by repeated interaction in massive eccentric binaries \citep[e.g.,][]{Kashi2013}.

\subsubsection{Gap transients: intermediate luminosity optical, or red transients}	
\label{sssec:ilots}

ILOTs/ILRTs such as SN~2008S \citep{Prieto2008}, NGC~300~2009OT-1 \citep{Bond2009}, or iPTF10fqs \citep{Kasliwal2011} at the luminous end of the gap have been interpreted as faint terminal explosions, because of the complete disappearance of the progenitor after the outburst. They are associated with dusty environments and are  tentatively hypothesised to derive from electron-capture supernovae \citep{Botticella2009} after a short and dusty transition phase lasting approximately 10\,000 years. \citet{Kashi2016} interpret ILRTs as less massive versions of supernova impostors and they argue that both groups are non-terminal outbursts due to mass accretion onto a companion in an eccentric orbit. As we already pointed out in Section~\ref{sssec:modelling-binary-interactions}, this rests on finding a suitable accretion model, which at the moment is beyond our understanding.

\subsubsection{Gap transients: luminous red novae}	
\label{sssec:lrn}

\begin{figure}
\centering
\includegraphics[width=0.4\textwidth]{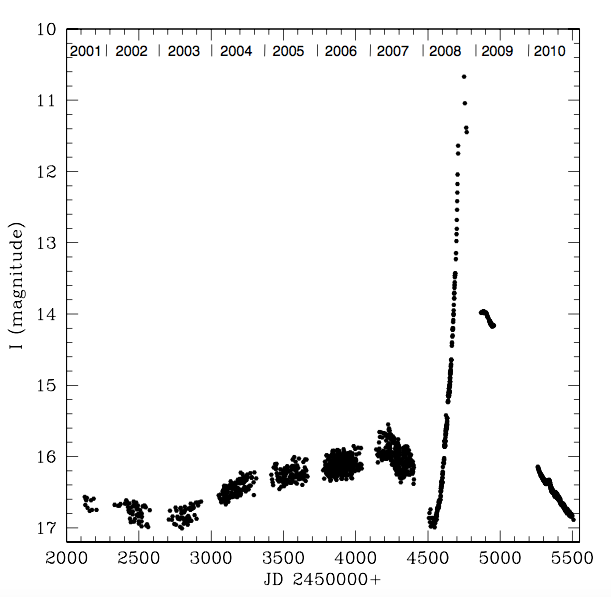}
\caption{The $I$ band light curve of the gap transient V1309~Sco $I$- from \citealt{Tylenda2011}, showing a slow rise in brightness over $\sim$4 years prior to the outburst (Section~\ref{sssec:lrn}). The range of brightness seen before the dip at JC24500004500 is due to variability caused by ellipsoidal distortion in the pre-outburst, contact binary. Due to the absence of the binary after the outburst this is the best observational example of a merger we have to date. {\it Credit: image from \citet{Tylenda2011}}}
\label{fig:v1309sco}
\end{figure}

LRNe are thought to be violent binary interactions \citep{Iben1992,Soker2003}. The best studied example \rgi{is} V1309~Sco \citep[][Fig.~\ref{fig:v1309sco}]{Tylenda2011}, a system \rgix{that was} discovered serendipitously, but which \rgi{is} in the OGLE field of view \rgi{hence has} a long baseline of pre-outburst observations. V~1309~Sco is very likely be a merger because the light curve before the outburst showed a contact binary, and this binary disappeared after the outburst. 

Another well studied example of an LRN is V~838~Mon \citep{Bond2003}, \rgi{which} was followed by several similar outbursts \citep[e.g.,][]{Williams2015}. More massive objects can be seen out to larger distances and as such it is likely that these more rare phenomena will be those observed more often. An example of a massive transient that likely shared many characteristics with V~838 Mon is M101-OT \citep{Blagorodnova2016}. This object peaked in brightness in 2014 and 2016. Archival photometry shows a binary system with a mass ratio of 0.9 and a total mass of 20~\msun\ that underwent a common envelope phase as the primary ended core hydrogen fusion. The mass of the progenitor fills the gap between the lower mass examples such as V~838~Mon (5-10~\msun) and the more massive examples such as NGC~4490-OT \citep[30~\msun;][]{Smith2016b}. A model of the scenario also predicts that the binary survived the common envelope phase. 

\begin{figure}
\centering
\includegraphics[width=0.48\textwidth]{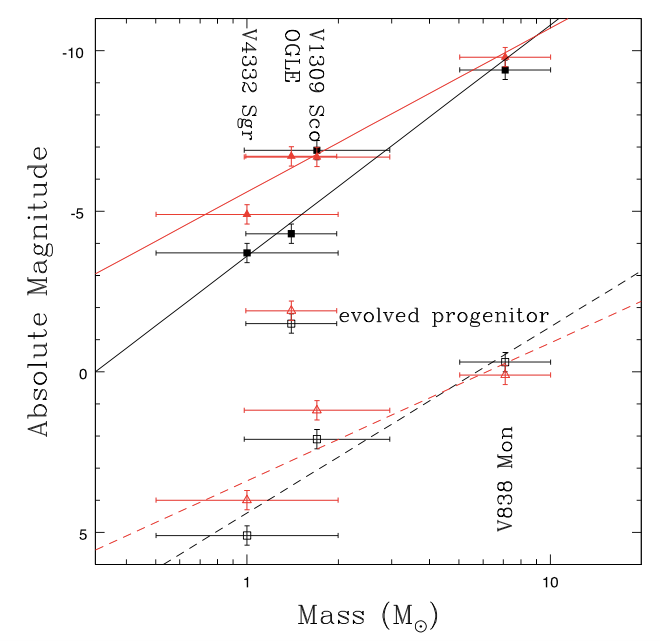}
\caption{Absolute magnitudes of the progenitors (open symbols) and transient peaks (filled symbols) in the $V$ (squares) and $I$ (triangles) bands, as a function of the progenitor mass estimates for gap transients (Section~\ref{ssec:intermediate-luminosity-optical-transients}). The best power-law fits are also shown. {\it Credit: adapted from figure 5 of \citet{Kochanek2014}. }
}\label{fig:kochanek14}
\end{figure}

\citet{Smith2016b} also point out that NGC~4490-OT fits the correlation between merger mass and peak luminosity discovered by \citet[][see Fig.~\ref{fig:kochanek14}]{Kochanek2014}, adding a more massive, more luminous data point. They also show that there could be a correlation between mass, peak luminosity and the duration of the outburst. Finally they point out how the light echo spectrum of $\eta$~Car \citep{Rest2012} is similar to the spectrum of NGC~4490-OT and that of V~838~Mon at some epoch, connecting LRNe and supernova impostors. They propose that $\eta$~Car could be an even more massive example within the same correlation, having had a brighter and longer-lasting outburst. 

\subsubsection{Gap transients: sundry}	
\label{sssec:sundry} 


Other gap transients not fitting well within the characteristics of the previous classes are the .Ia supernovae, first conjectured by \citet[][see also \rgi{\citealt{Shen2010}}]{Bildsten2007} to be surface detonations on CO WDs following accretion from a less massive, companion WD. The best case of such a supernova detected to date was described by \citet{Kasliwal2010}. 


Another type of gap transients are the "calcium-rich gap transients". Like type Ia supernovae they have no hydrogen, but they tend to be 10-30 times fainter. They have very high calcium abundances, as inferred from their nebular phase spectra. Current (small) samples place them in the outskirts of galaxies \citep{Kasliwal2012b}. Theories of their formation abound, but each has at least one serious flaw. For example \citet{Perets2010} suggests that calcium-rich gap transient SN~2005E was a helium detonation on a WD accreting from a helium WD. Such sub-Chandrasehkar detonation models \citep{Woosley2011} also do not reproduce the light curve. These transients were also explained as the tidal detonation of a low mass WD, which could produce some of the calcium \citep{Sell2015}: an intermediate-mass black hole passing by the WD in dense cluster environments could trigger the detonation; alternatively the black hole could be in orbit with the WD in a triple system where a wider co!
 mpanion tightens the inner binary.

\subsection{Radio transients}
\label{ssec:radio transients}

The \rgi{radio-transient} sky is still largely unexplored. Fast radio transients are intense, millisecond bursts \rgi{of} uncertain origin, so far detected at 1.2~GHz and 1.6~GHz. Their large dispersion measures and high galactic latitudes suggest that they have a cosmological origin \citep{Lorimer2007}. They are not associated with any known astrophysical object, but candidates include pulsar-planet binaries \citep{Mottez2014}, binary WDs \citep{Kashiyama2013}, pulsars and magnetars \citep{Petroff2015}. There are also predictions that neutron star mergers forming a neutron star with a mass larger than the non-rotating limit, may eventually spin down and collapse to form a black hole. As their field lines cross the newly formed horizon they snap and the resulting outwordly propagating magnetic shock dissipates as a short radio burst \citep[e.g.,][known as the "blitzar" model]{Ravi2014}. In a different model, neutron star mergers would emit a fast radio burst just before they!
  coalesce, when their magnetic fields become synchronised with the binary rotation \citep{Totani2013}. If neutron star mergers did indeed produce fast radio transients, then the neutron star-neutron star merger rate should be at the high end of the range predicted \citep{Abadie2010}.

Slow radio transients might include instead supernovae and binary neutron star mergers, as well as tidal disruption of stars by supermassive black holes \rgi{which, while not directly related to binary evolution,} shed light on disks and jets (cf.~Section~\ref{ssec:jets}). Surveys \rgi{of slow transients} are planned with a range of instruments \citep[][]{Caleb2016}. The upcoming ASKAP will likely add vital evidence to what is already known by X-ray transient surveys \citep{Macquart2014,Donnarumma2015}.

\subsection{Gravitational Wave Sources}
\label{ssec:gravitational-wave-sources}
\begin{figure*}
\centering
\includegraphics[width=0.48\textwidth]{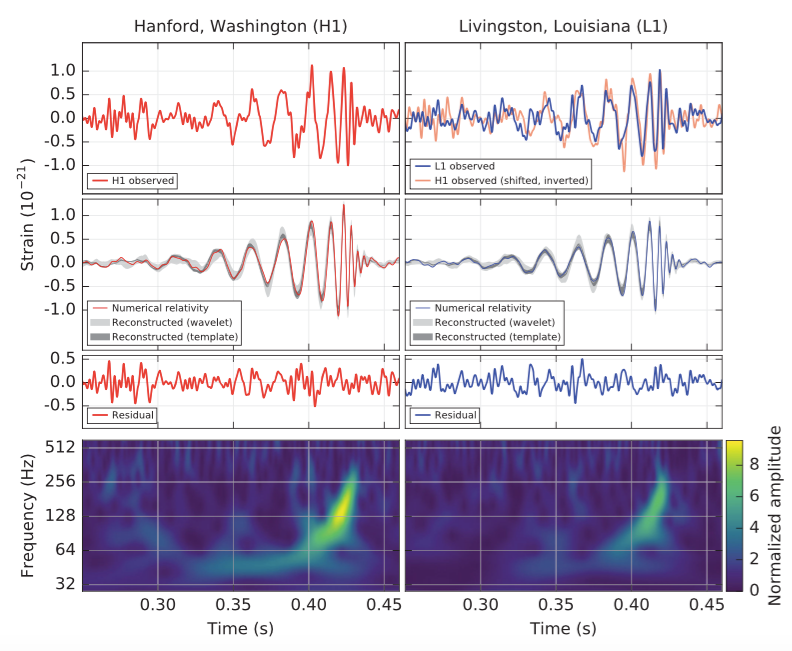}
\includegraphics[width=0.51\textwidth]{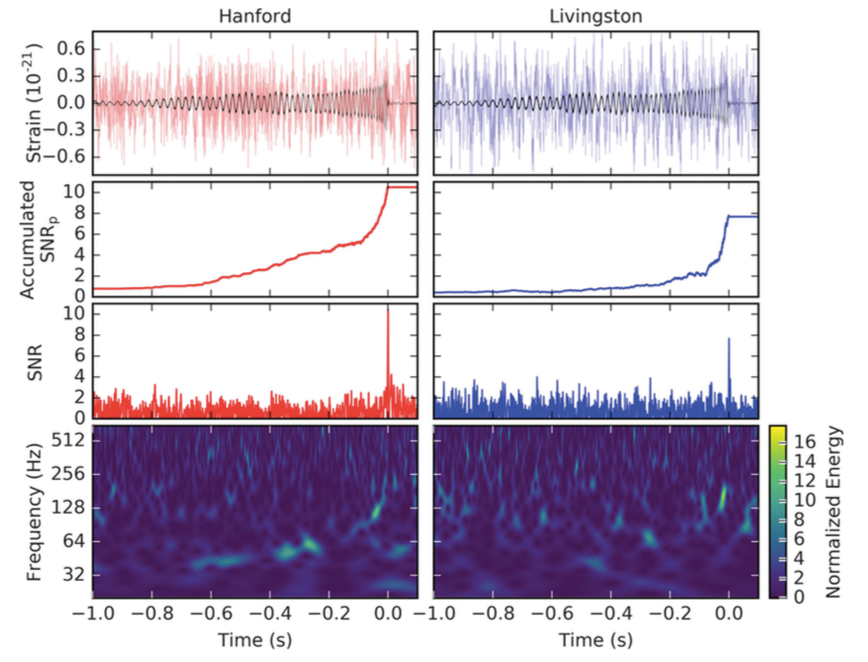}
\caption{The gravitational-wave events (Section~\ref{ssec:gravitational-wave-sources}) of the only two confirmed detections so far: GW150914 (left panel - figure from \citet{Abbott2016}) and GW151226 (right panel) observed by the LIGO Hanford and Livingston detectors. Left panel: times are shown relative to September 14, 2015 at 09:50:45 UTC. Top row, left: H1 strain. Top row, right: L1 strain. GW150914 arrived first at L1 and 6.9~ms later at H1; for a visual comparison, the H1 data are also shown, shifted in time by this amount and inverted (to account for the detectors' relative orientations). Second row: Gravitational-wave strain projected onto each detector in the 35-350~Hz band. Solid lines show a numerical relativity waveform for a system with parameters consistent with those recovered from GW150914. Shaded areas show 90 per cent credible regions for two independent waveform reconstructions. Third row: Residuals after subtracting the filtered numerical relativity wavef!
 orm from the filtered detector time series. Bottom row: A time-frequency representation of the strain data, showing the signal frequency increasing over time. Right panel: Times are relative to December 26, 2015 at 03:38:53.648 UTC. First row: Strain data from the two detectors. Also shown (black line) is the best-match template from a non-precessing spin waveform model. Second row: The accumulated peak signal-to-noise ratio as a function of time when integrating from the start of the best-match template, corresponding to a gravitational-wave frequency of 30 Hz, up to its merger time. Third row: Signal-to-noise ratio time series. Fourth row: Time-frequency representation of the strain data around the time of GW151226. In contrast to GW150914, the signal is not easily visible. {\it Credit: figure 1 of \citet{Abbott2016} and figure 1 of \citet{Abbott2016f}.}}
\label{fig:abbott16-chirp}
\end{figure*}

The historical detection of gravitational waves by the {\it Advanced Laser Interferometer and Gravitational Wave Observatory}, LIGO, was interpreted as the merging of a binary black hole \citep{Abbott2016}. Aside from providing a test of general relativity exactly a century after its formulation \citep{Einstein1916}, this detection has opened a new window on the study of binary stars. A phenomenal amount of information has been\rgi{, and remains to be,} derived from these detections \citep{Abbott2016b}. Crucially, this discovery proves the existence of a type of binary that was \rgi{previously} hypothetical. 

A gravitational wave passing LIGO alters the differential length, $L$, of the interferometer's perpendicular arms so that the measured difference  is $\Delta L(t) = \delta L_x - \delta L_y =  h(t)L$, where $L = L_x = L_y$ and $h$ is the gravitational-wave strain amplitude projected onto the detector. The first detection took place on the 19$^{\rm th}$ September 2015 and was truly bright with a strain amplitude of $1.0 \times 10^{-21}$ (Fig.~\ref{fig:abbott16-chirp}, left panel). \rgi{The two} black holes were deduced to have masses of $36^{+5}_{-4}$~\msun\ and $29\pm4$~\msun, while the final black hole mass was determined to be $62\pm4$~\msun; $3.0\pm0.5$~\msun~$c^2$ was radiated in gravitational waves, with a peak gravitational wave luminosity of $3.6^{+0.5}_{-0.4} \times 10^{56}$~erg~s$^{-1}$ and a luminosity distance of 410$^{+160}_{-180}$~Mpc. The merger must have formed at low metallicity or else the masses of the two black holes would have been decreased by stellar win!
 ds. It is still not clear whether the merger was a binary coalescence or resulted from a dynamic encounter in young or old dense stellar environment. They either formed at low redshift and merged promptly, or formed at higher redshift but took several gigayears to merge.

A second signal was detected on the 26$^{\rm th}$ December 2015 (Fig.~\ref{fig:abbott16-chirp}, right panel). The "Boxing Day" event was interpreted as the merger of two black holes with initial masses $14.2_{-3.7}^{+8.3}$~\msun\  and $7. 5\pm 2.3$~\msun, and a final black hole mass of $20.8_{-1.7}^{+6.1}$~\msun\ \citep{Abbott2016d}. This detection had a strain amplitude of $3.4^{+0.7}_{-0.9} \times  10^{-22}$, smaller than GW150914, and the signal was spread over a longer time interval. The source had a peak luminosity of $3.3^{+0.8}_{-1.6} \times 10^{56}$~erg~s$^{-1}$, a luminosity distance of $440^{+180}_{-190}$~Mpc and a source redshift of $0.09^{+0.03}_{-0.04}$.

A third signal was too faint to be classified as a detection and was instead named LTV151012 \citep{Abbott2016d}. Thus the observing run that took place between 12$^{\rm th}$ September 2015 and 19$^{\rm th}$ January 2016 detected two events in the total mass range 4-100~\msun\ \citep{Abbott2016d}.

\begin{figure}
\centering
\includegraphics[width=0.48\textwidth]{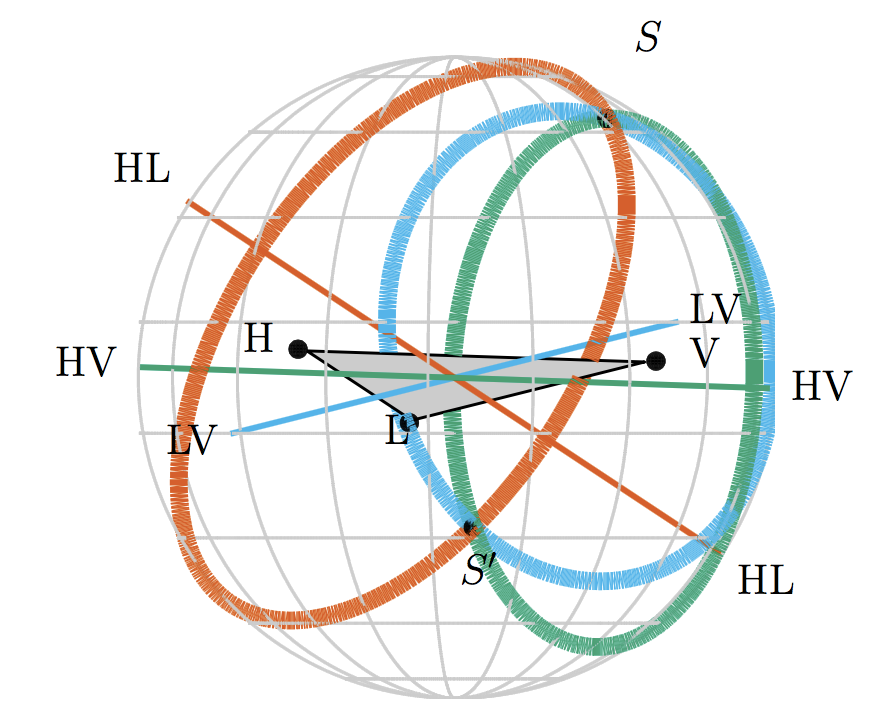}
\caption{Gravitational wave source localisation (Section~\ref{ssec:gravitational-wave-sources}) by triangulation possible for the 3-detector aLIGO-AdV network
The three detectors are indicated by black dots, with LIGO Hanford labeled H, LIGO Livingston as L, and Virgo as V. The locus of constant time delay (with associated timing uncertainty) between two detectors forms an annulus on the sky concentric about the baseline between the two sites (labeled by the two detectors). For three detectors, these annuli may intersect in two locations, one coincident with the true source location (S), while the other (S') is its mirror image with respect to the geometrical plane passing through the three sites. A precise localisation is key for follow up observations that seek to identify an electromagnetic signature. {\it Credit: image adapted from figure 4 of \citet{Abbott2016c}}.}
\label{fig:abbott16c}
\end{figure}

The ability to locate the detected signal to within reasonable areas of the sky (5-20~sq. deg.) is crucial to hunt for electromagnetic counterparts to the gravitational wave source. For this we must wait for an additional detector with a sensitivity within a factor of two of the other two (Fig.~\ref{fig:abbott16c}). At present it is likely that a third LIGO will be constructed in India \citep{Abbott2016c}.

It is  likely that gravitational waves from neutron stars mergers will \rgi{soon} be detected. \rgi{Observed m}erger rates of neutron stars and black holes will impose new constraints on the physics of \rgix{the} binary \rgi{interactions that precede} the merger, including the elusive common envelope interaction (Section~\ref{sssec:modelling-binary-interactions}). 
\rgi{Forecasts} of the LIGO-observable merger rates \rgi{range} between 0.04 and 400 events per year \citep{Abadie2010}. It is likely that the rate is closer to the higher end of estimate range, considering the first detection took place soon after the start of the operations of Advanced LIGO.
If an afterglow were to be detected \citep{Loeb2016}, additional properties such as redshift could open novel tests of cosmology.

Bursts of gravitational waves shorter than 1 second in duration are predicted from core-collapse supernovae \citep{Ott2009}, neutron stars collapsing to black holes \citep{Baiotti2007}, cosmic string cusps \citep{Damour2001}, star-quakes in magnetars \citep{Mereghetti2008}, pulsar glitches \citep{Andersson2001}, and signals associated with gamma ray bursts \citep{Abadie2012}. We also could expect a gravitational waves signal from sources emitting over long periods of up to hundreds of seconds and most likely associated with non symmetric hydrodynamic instabilities predicted to occur immediately following the formation of a neutron star in a core-collapse supernova \citep{Abbott2016d}.

More importantly, there is no telling what surprises might lurk in this categorically new type of astronomical data which, unless we have been exceedingly lucky, will be plentiful from LIGO, Virgo and the soon to be built Indian counterpart. 


\section{\rgi{CONCLUSIONS}}
\label{sec:concluding-remarks}

The governing principles of stars were identified in the \rgi{first half of the 20th century}, when the source of their \rgi{longevity} was determined to be nuclear \rgi{energy}. Since then the understanding of stars has gained in depth and sophistication. As the field of stellar astrophysics matured, it {increasingly took on} a service role \rgi{for other fields of astrophysics}. For example, in order to determine how our Galaxy \rgi{formed,}  "galactic archeologists" need to track the point and time of origin of millions of stars \citep[e.g.,][]{Martell2016}. To do this they need precision kinematics and abundances of every star, something that has propelled forward studies of stellar structure and resulted in further improvements in the modelling of stellar interiors and photospheres.

In the context of stellar structure and evolution, the effects of companions were typically either not thought severe enough to alter the course of stellar evolution, or they were observed to be so severe to move the star into a class of its own, interesting only to a few scientists. Stellar astrophysicists have been and still are preoccupied with a series of complexities governing processes in single stars, and it is understandable that the effect of binarity be set aside not to complicate matters beyond the point when the system cannot be modelled.

It is only natural that interactions with companions can significantly impact the future evolution of a star. What has been missing to make connections between phenomena and duplicity more concrete are good statistics of the binary fraction, period and mass ratio distributions, a knowledge that has recently improved. Today it is clearer that to interpret observations of stars at any evolutionary phase, we must entertain the possibility that an interaction has taken place.  Massive stars in particular often interact with companions that can therefore influence every stage of their lives, particularly the phases where they become giants and lose copious mass.  Massive stars cause cosmologically detectable outbursts and are key players in injecting energy and momentum into their environments, something that drives galactic evolution. This means that we must strive to include these interactions in theories of massive star evolution.

These realisations have driven an increase of observational platforms (telescopes and surveys). Theoretical codes and methods that have been the pillars of stellar structure and evolution studies have been developed further to include binary interactions, alongside new codes and methods.  Binary studies are becoming important not only in stellar evolution, but in a range of other fields in \rgi{astrophysics}, such as the study of jets, applicable to star and planet formation as well as active galactic nuclei, or the production of gravitational waves. 

This underscores the importance of an improved theoretical framework to interpret the large amount of observations that are already  accumulating. In particular, some of the key phenomena are those grouped under the heading of mass transfer. They are extremely complex and only a great improvement of our modelling capabilities will be able to match observations. Three-dimensional modelling is likely necessary, but adding the necessary complexity is still unfeasible. However algorithms and computer power are both improving dramatically and with it will 3D simulations. As the backbone of simulations' machinery is improved, a concerted effort to keep all branches of binary research working closely alongside is necessary. 

One hundred years after Eddington figured out how stars work, we are adding a new ingredient to stellar evolution. While increasing complexity, the inclusion of binary interactions also adds clarity, because now, new and old stellar phenomena have a chance to find an explanation within an expanded stellar evolution paradigm.

\acknowledgements
We are grateful to a number of colleagues, including an anonymous referee, who have commented on this manuscript, something that has improved this review. In particular we thank: Igor Andreoni, Geoff Clayton, JJ Eldridge, Adam Jermyn, Mansi Kasliwal, Tony Moffat, Daniel Price, Ashley Ruiter, Nathan Smith and Noam Soker. OD would like to thank the Australian Research Council's Future Fellowship Programme via grant FT120100452. RGI thanks the UK Science, Technology and Facilities Council for supporting his Rutherford Fellowship, grant number ST/L003910/1. This research has made use of NASA's Astrophysics Data System Service.

\bibliographystyle{apj}
\bibliography{bibliography}

\end{document}